\documentclass[final,5p,twocolumn]{elsarticle}

\usepackage{amssymb} 
\usepackage{enumitem} 
\usepackage{dblfloatfix} 

\journal{Information and Software Technology}

\begin{document}

\begin{frontmatter}


\title{Collaborative Software Design and Modeling in Virtual Reality}

\cortext[cor1]{Corresponding author.}

\author[1]{Martin Stancek}
\author[2,1]{Ivan Polasek\corref{cor1}} \ead{ivan.polasek@fmph.uniba.sk}
\author[1]{Tibor Zalabai}
\author[2]{Juraj Vincur}
\author[3]{Rodi Jolak} 
\author[4]{Michel Chaudron}

\affiliation[1]{organization={Gratex International a.s.},
            addressline={Galvaniho 17/C}, 
            city={Bratislava },
            postcode={821 04}, 
            country={Slovakia}}
            
\affiliation[2]{organization={Comenius University, Faculty of Mathematics, Physics and Informatics},
            addressline={Mlynská dolina F1}, 
            city={Bratislava},
            postcode={842 48}, 
            country={Slovakia}}
            
\affiliation[3]{organization={RISE Research Institutes of Sweden},
            addressline={Box 857}, 
            city={Borås},
            postcode={501 15}, 
            country={Sweden}}
            
\affiliation[4]{organization={Eindhoven University of Technology},
            addressline={O.L. Vrouwestraat 1}, 
            city={Eindhoven},
            postcode={5612 AW}, 
            country={Netherlands}}

\begin{abstract}
\textbf{Context: }Software engineering is becoming more and more distributed. Developers and other stakeholders are often located in different locations, departments, and countries and operating within different time zones. Most online software design and modeling tools are not adequate for distributed collaboration since they do not support awareness and lack features for effective communication. 
\\ \textbf{Objective: }The aim of our research is to support distributed software design activities in Virtual Reality (VR).
\\ \textbf{Method: }Using design science research methodology, we design and evaluate a tool for collaborative design in VR. 
We evaluate the collaboration efficiency and recall of design information when using the VR software design environment compared to a non-VR software design environment. Moreover, we collect the perceptions and preferences of users to explore the opportunities and challenges that were incurred by using the VR software design environment. 
\\ \textbf{Results: }We find that there is no significant difference in the efficiency and recall of design information when using the VR compared to the non-VR environment. Furthermore, we find that developers are more satisfied with collaboration in VR. 
\\ \textbf{Conclusion: }The results of our research and similar studies show that working in VR is not yet faster or more efficient than working on standard desktops. It is very important to improve the interface in VR (gestures with haptics, keyboard and voice input), as confirmed by the difference in results between the first and second evaluation.
\end{abstract}




\end{frontmatter}


\section{Introduction}\label{sec2}

The engineering of software systems is a socio-technical process where software developers and other stakeholders having different expertise and experiences are involved \cite{bib74}. One of the fundamental phases of software engineering is software design where important aspects of the problem and solution domain are explored, discussed, and often described in software models \cite{bib75}.

Whiteboards are widely used by developers to design and create software system models \cite{bib76}. However, software engineering, including software design and modeling, is becoming more and more distributed \cite{bib77}. Indeed, teams are often located in various departments, divisions, countries, and operate within different time zones \cite{bib78}. 

Most online design and modeling tools do not provide immersive collaboration environments for collaborative developers \cite{bib77}. Indeed, these tools lack features addressing collaboration awareness\footnote{Collaboration Awareness: refers to the awareness of the activities of the collaborators. Furthermore, it provides context and information on interactions with the shared workspace \cite{bib47}.} and effective communication, which could influence the effectiveness of the design process and thus the overall quality of the software.

Nowadays, Virtual Reality (VR) is an attractive and accessible technology that is used to build VR-based environments. These VR environments allow team members to work together in a shared virtual space, regardless of their physical location, which can increase flexibility \cite{bib31}. In addition, VR environments provide an immersive telepresence experience and support awareness in remote collaboration.

At the beginning of this research there was a debate with the developers at the software company Gratex International, who had had to collaborate with their colleagues in Seoul and Australia on insurance software and data models for contracts, etc. They asked us if the decisions over model changes and the design of new modules would not be more intense and natural if we were also able to see what the others are looking at in VR and what they are doing (differentiation by a unique colour for each person). Many colleagues felt a disconnect from the other team members in Seoul. We needed to somehow overcome the long distance between Europe, Australia and Seoul and get closer in VR. It was also about saving costs, so that other colleagues wouldn't have to travel (flights and staying in Seoul and Australia are not cheap) and waste time on planes and in airports if they had another project here in Europe.

Later came the change in work due to the pandemic, as we needed to collaborate without endangering our health. Subsequently, after the initial waves of the pandemic and the restrictions, there was a greater preference for working from home among employees, and utilizing VR for designing or presenting models became an intriguing alternative.

We were also interested in VR as a means of changing our working style, in addition to the constant sitting behind a standard computer. We were curious about whether immersion in VR for collaborative modeling could improve the results or, at the very least, enhance the designers' experience.

Several arguments for the use of VR have emerged:
\begin{itemize}
\item Overcoming the geographical distance of distributed teams and reducing travel and accommodation costs.
\item Providing solutions for online work within a home office, especially during situations like illness or pandemics.
\item Immersion in collaborative modeling by surrounding ourselves with models of the system.
\item Changing the work style and breaking the monotony of sitting at the computer.
\end{itemize}

Brettschuh et al. in \cite{Y12} stated that VR tools could be one of the promising implementations of Industry 4.0. They also reported the main domains of VR closely related to our area, which include Product design, Process design and visualization, Collaboration and Co-creation, Product Presentation, and Prototyping. The majority of studies they reviewed utilized the effect of immersion for their own purposes, leveraging VR to gain various advantages.

Berg and Vance argue in \cite{Z11} that VR supports a sense of team engagement, which leads to better discussions and increased participation of team members in the decision-making process.
They demonstrated in their survey \cite{Z12} that VR has been accepted and actively used in numerous industries to support decision-making and enable innovation.
VR can involve end-users in the design process better by providing different alternatives.

Due to the impact of the pandemic situation, VR has been considered a potential answer to online teaching for software engineering and learning enhancement \cite{3APSEC}. In \cite{3APSEC}, they found that VR improves the experience in software engineering presentations, and the students of SE feel more presence and have better interactions during the presentations in VR.
Borst et al. in \cite{555} reported that VR could increase motivation and engagement. R. Oberhauser in \cite{bib20} sees various positive indicators that VR-UML can show advantages where more complex and multi-diagram models are involved (and by inference hypermodeling). He anticipates that the immersive experience of UML models in VR adds qualitative aspects that users prefer.

Romano et al. in \cite{1Romano} observed that the use of immersive virtual reality leads to higher satisfaction among the participants. Their findings suggest that virtual reality might be a viable tool in software visualization, and their experiment seems to justify further research on its use in this area.

The aim of this research is to investigate the use of VR as a means to support as well as improve the efficiency of distributed software design activities. To achieve that, we employ the design science research methodology \cite{bib17} to design and evaluate a VR software design environment that enhances the collaboration experience between remote or geographically distributed software developers.

In particular, we address the following research questions:
\begin{itemize}
   
    \item \textbf{RQ1.} \emph{What is the impact of the created VR environment on the efficiency of collaborative, distributed software development?}

    We evaluate the collaboration efficiency when using the VR software design environment compared to a non-VR software design environment.
    
    \item \textbf{RQ2.} \emph{What is the impact of the created VR environment on the recall of design information during collaborative, distributed software development?}

    We evaluate the recall of design information when using the VR software design environment compared to a non-VR software design environment.
    
    \item \textbf{RQ3.} \emph{What are the perceptions and preferences of users when using the VR-based compared to the non-VR software design environment?}
    
    We collect the perceptions and preferences of users to explore the opportunities and challenges that were incurred by using VR software design environments. These perceptions will be used to further improve the created environment in order to support the design activities of distributed software developers better. 
\end{itemize}

\subsection{Structure of the Paper}

The background and related work are presented in Chapter~\ref{sec1}. The approach that includes the design and creation of the VR environment is described in Chapter~\ref{sec3}. The evaluation of the VR environment and experiments are presented and discussed in Chapter~\ref{sec4}. Finally, concluding remarks and future work are presented in Chapter~\ref{sec7}.

\section{Related Work}\label{sec1}

\subsection{Software design, modeling and collaboration}

There are several methods that aim to simplify complex and long-term software development processes. Several methods are known for projecting UML into three-dimensional space because large 2D diagrams might be hard to read and understand. Casey and Exton proposed a new method \cite{bib2}, where the elements in the diagrams are geons. Geon is basically a simple three-dimensional nonstandard UML element, e.g. square, pyramid or sphere. 

Other approaches use the advantages of 3D space and layered 2.5D diagrams \cite{bib11} or virtual reality \cite{bib73}. The Ferenc prototype \cite{bib4} supports these layered diagrams in web applications with real-time collaboration and visualization of the user's editing history. The conversion of diagrams from 2D to 3D followed by visualization in augmented or virtual reality is researched in SAR process \cite{bib81} or VisAr3D \cite{bib8}. They are focusing on large-scale systems and experimenting with virtual and augmented reality in relation to education. 

There are papers dealing with collaborative modeling drawing attention to the importance of the human element and its support \cite{bib68, bib67}. A different way of collaborative modeling is OctoUML \cite{bib1} which uses sketchy diagrams and transforming them into formal UML notation. OctoUML supports a lot of inputs, e.g. interactive whiteboards. Interaction between a UML model and a human was researched by Anders Mikkelsen et al. \cite{bib5}. They studied the interaction in augmented reality on Microsoft Hololens devices. They tried to make UML simpler by interaction, not by changing the UML itself. 

Fuks et al. \cite{bib46} adopted the 3C model (coordination, communication, and cooperation) and declared that awareness mediates and fosters all three aspects of collaboration. We can use this model as a base for analyzing and designing groupware. 

Dourish and Belloti \cite{bib47} defined awareness as “an understanding of the activities of others, which provides context for one’s own activities.” Gutwin and Greenberg \cite{bib48} claimed that every collaborator should be intuitively aware of current related aspects, such as who is in the workspace, where they are located, what they are working on, as well as past related aspects, such as how this artifact came to be in this state or who made this change and when. They confirmed the importance of user’s awareness of co-workers in a shared workspace together with other studies \cite{bib49, bib50}. 
They declared that good awareness provides the following benefits:
\begin{enumerate}
    \item collaborators will not miss a chance to collaborate,
    \item minimizing the interruptions of co-workers at inappropriate time,
    \item better contextual understanding of where assistance is required,
    \item elimination of unnecessary need for communication,
    \item prediction of the others’ actions and therefore easier decision on choosing their next task,
    \item work redundancy is eliminated and division of labor is simplified.
\end{enumerate}

Based on the above, we can declare that awareness of co-workers and their activities increase work efficiency and productivity in multi-user workspace. However, this does not solve the problem of readability of complex and large-scale UML models of software systems. Many research papers propose that modeling and visualizing a system in 3D space can eliminate this problem and introduce improvements.

The importance of collaboration between stakeholders is also confirmed by the observation of project failures in \cite{bib60}. The six most common reasons include incomplete requirements, lack of user participation, lack of resources, unrealistic expectations, lack of executive support, and change in requirements and specifications. None of them deals with programming languages, technological environments, or hardware components. Five of the six mentioned reasons are due to the interaction. Collaboration and communication in the development of the software project are important for the execution of successful and effective projects.

Franzago et al. in \cite{bib61} classified existing approaches of collaborative Model-Driven Software Engineering (MDSE) from 106 selected papers clustered into 48 studies. 

They summarized the most important characteristics of collaborative tools: 
\begin{itemize}
    \item modeling language (mostly UML), 
    \item editor (for CRUD modifications and operations with model elements), 
    \item application domain (mobile, web or business application, and orthogonally: user interface, business logic, data layer, etc.), 
    \item multi-view support (using decomposition to reduce complexity), 
    \item versioning support, 
    \item conflict management (mainly conflict detection or locking elements to prevent conflicts), communication support, 
    \item collaboration type (Synchronous in real-time or Asynchronous work on the same artifacts but offline or not at the same time).
\end{itemize}

Yigitbas et al. \cite{bib37} claimed that standard modeling applications (Lucidchart \cite{bib42}, GenMyModel \cite{bib41}, etc.) containing remote collaboration on a 2D UML notation do not support a natural way of teamwork comparable to editing a model on a whiteboard together in one room.

Lucidchart \cite{bib42} supports UML and Business Process Model Notation (BPMN). GenMyModel \cite{bib41} supports UML and users can see where their co-workers are editing the model but do not have voice chat to communicate with them. Another tool is the collaborative learning environment for UML modeling CoLeMo \cite{bib40} designed for participants studying UML modeling. It can be used as a platform for collaborative software design. Teaching software engineering in Virtual reality was proposed by Herpich et al. \cite{bib43} in their prototype Virtual Lab through using panels embedded into the 3D virtual environment. Each panel represents one instance of the GenMyModel modeling tool.

\subsection{VR tools supporting software modelling}

\begin{figure}[t]
    \centering
    \includegraphics[width=0.48\textwidth]{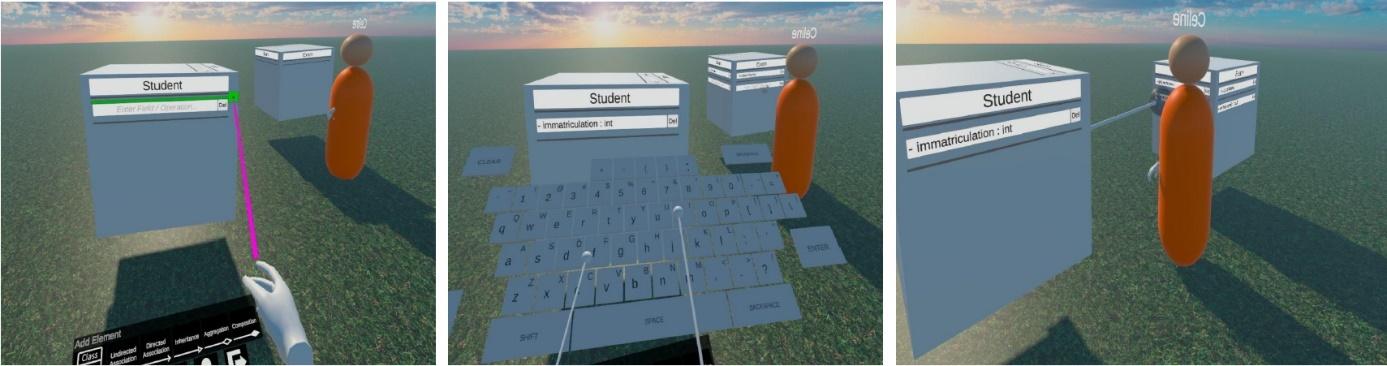}
    \caption{Collaborative VR modeling environment \cite{bib37}}
    \label{fig:colab_vr}
\end{figure}

Collaborative modeling in virtual reality is proposed in \cite{bib37}. The authors believe that the use of 3D cubes (Figure~\ref{fig:colab_vr}) and tubes instead of 2D class rectangles and relationship lines could result in a more natural experience for users because the real world consists only of 3D objects. Every cube side shows the same 2D Class notation (name, attributes, and methods) similar to 2D UML. They built a prototype with these cubes on an unlimited planar grass field under a blue sky to depict a pleasant real-world environment for the user to walk. Visualization of the cubes is similar to visualizations of BPMN in 3D or VR \cite{bib39}.

It seems that trivial presentation of the boxes on a flat terrain does not contribute to reducing complexity nor promoting comprehensibility. It will also be probably difficult to get an overview of where other collaborators are located. Maybe the unused 3D space in the air would help with a very complex diagram to reduce the number of edge intersections and make it easier to layout diagrams or create the clusters. Authors could also use the top side of the cube for additional information (stereotype, state, module, etc.). 

In the evaluation process in \cite{bib37} the participants had to create one class diagram in VR and another one in the web application (Lucidchart). On average, participants needed more time in VR than in the Web application to complete their tasks (13:42 versus 07:35). Also, the number of the errors per minute was higher in VR. Better VR results were achieved in subjective evaluation (Interactivity, Co-Presence). “13\% (3 out of 24) wanted to use the VR app over the web application and 37\% (9 out of 24) vice versa. However, half of the participants indicated that they would like to use both applications, each for specific situations '\cite{bib37}.

Augmented reality with MS Hololens was tested in \cite{bib44} for students to make UML more accessible.  Zenner et al.  \cite{bib38} proposed a system that visualizes EPC (Event-Driven Process Chain) models in VR 3D. Their approach focused on model understandability and user experience by comparing the HMD VR experience with viewing a 2D process model. They observed that users understand the model faster in 2D (objective effectiveness and efficiency) while the users interest and immersion is higher in VR.

In augmented reality around the desktop, we can visualize other diagrams \cite{bib26} behind the primary one for additional information (e.g. authors, element types, tags, metrics, bugs, design patterns, bad smells, etc.). The same approach is possible in VR behind or around our whiteboards.

In software engineering monitors with a mouse and keyboard are mostly used, which may limit natural perception and creativity of users \cite{bib21}. Therefore, it might be very useful to use virtual reality as a tool for software development \cite{bib3}. In the 3D space of VR, we can see the distance and scale of objects better than on standard monitors. VR technologies are now more affordable, which also creates new opportunities in the field of software design \cite{bib59}. The following advantage of virtual reality is the indirect effect of movement on human creativity. Movement, for example, walking, has a positive effect on thinking, so it can increase the effectiveness in software development \cite{bib6}. In real life, there is a possibility to determine indoor location of humans researched by Arjan J.H. Peddemors et al. \cite{bib7}. They used Bluetooth technology of the mobile phone to determine where people are located in the building. They found out that the presence of people could be detected with room level granularity. We can use this approach in our prototype later, to determine the presence of the user's location.

\subsection{Discussion} \label{sec2_3}

\begin{table}[b]
    \centering
    \includegraphics[width=0.48\textwidth]{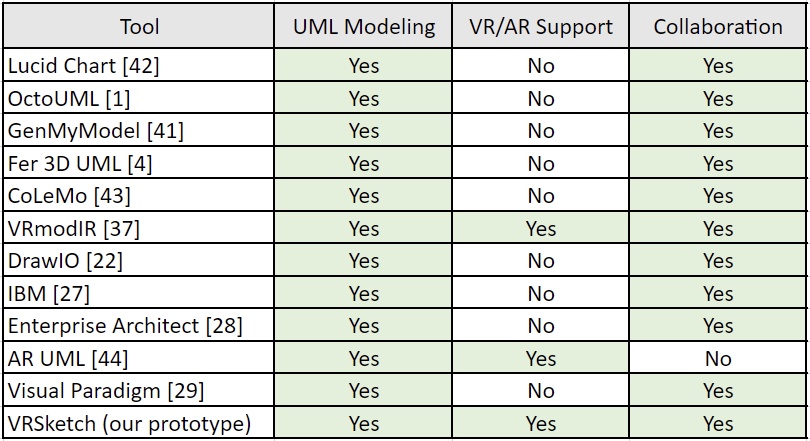}
    \caption{Modeling tools and their selected features (UML models editors, VR, and Collaboration)}
    \label{tab:modeling_tools}
\end{table}

At the end of our effort to create a short and concise overview of similar modeling tools and approaches, we can state that in addition to standard industrial CASE systems such as IBM Rational Software Architect \cite{bib27}, SPARX Enterprise Architect \cite{bib28} or Visual Paradigm \cite{bib29}, there are also academic experiments and concepts focused on modeling, collaboration, 3D space and VR utilization, etc. 

In the field of VR modeling tools, we found VmodlR as a comparable project, which also solves the enhancement of the immersion and satisfaction with collaboration using VR, but its approach is a little bit different: analysts and designers work on one model in VR 3D space, while we use 3D space for a dynamic and variable arrangement of whiteboards which developers are used to work on in their offices and laboratories. The advantage of whiteboards is also the possibility of distributing parts or modules into them as layers and their interconnections in this 3D space (see chapters 3.2 and 3.3). With 3D elements (cubes) on green grass under a blue sky in VmodlR \cite{bib37}, you cannot deal with this possibility of layering the model for reasonable and rational usage of 3D space.

At the end of the chapter, we created an overview (Table~\ref{tab:modeling_tools}) for a short comparative analysis of particular tools, where we can find two methods for collaborative modeling in virtual reality.

\section{Our Approach of Collaborative Software Design in Virtual Reality}\label{sec3}

We employ the design science research methodology (DSRM) \cite{bib17} to create and evaluate a VR software design environment. DSRM is an iterative process that involves problem exploration, the creation of a solution, the evaluation of the solution, and adaptation. 

The request for research and development from Gratex International was whether we are already able to achieve quality and speed in collaborative modelling in VR comparable to the classic desktop tool and whether it would speed up (or at least not slow down) the creation of models and achieve a better feeling of collaboration and encourage more frequent collaboration.

\begin{figure}[b]
    \centering
    \includegraphics[width=0.48\textwidth]{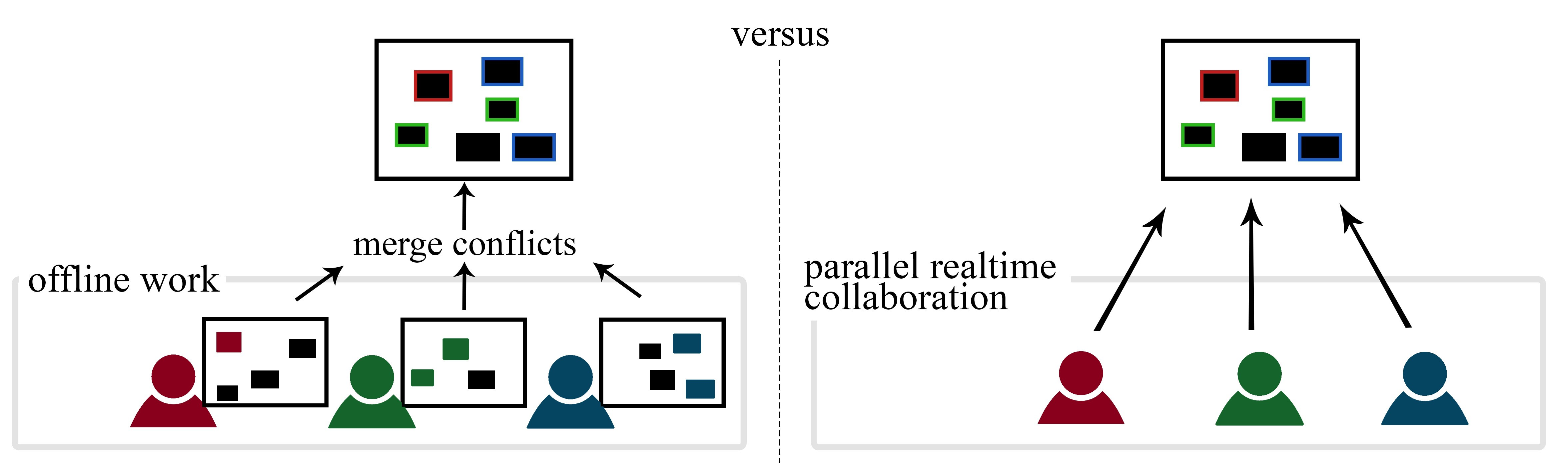}
    \caption{Results of merging and conflict resolution in offline work versus parallel real-time collaboration}
    \label{fig:conflict_diagram}
\end{figure}

\begin{figure}[h]
    \centering
    \includegraphics[width=0.47\textwidth]{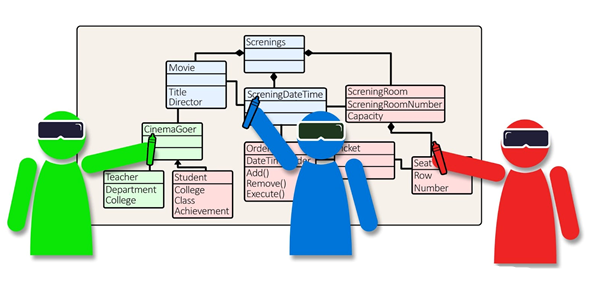}
    \caption{Collaborative modeling of software structure in virtual reality}
    \label{fig:structure_colaboration}
\end{figure}

Another idea of our approach is to avoid the problems with conflicts and merging of the results from the work offline (Figure~\ref{fig:conflict_diagram}) and support parallel collaborative model-driven development and collaboration of software development teams in virtual reality (Figure~\ref{fig:structure_colaboration}) to reduce communication gap between distributed developers working on software models. The goal is to increase user and team productivity and reduce the costs of remote software development. We designed a new tool in VR that supports sketching of the software structure in the Unified Modeling Language (UML) \cite{bib45} as standardized and widely used for the creation of software models that describe the architecture and functionality of the created system. 

According to related work in Chapter 2 we rely on the following assumptions and hypotheses, which we would like to prove/test in experiments with our tool:

\begin{enumerate}
    \item developers might not work worse or slower in virtual reality than on standard desktop systems, 
    \item developers might be more immersed in VR into their work and their models than on standard desktop systems and they might remember more of the model content from VR,
    \item developers might have a better feeling of collaboration because they will see the presence and materialized statures, movement, and gestures of their remote colleagues through the avatars in VR.
\end{enumerate}

\subsection{Concept} \label{sec3_1}

From the main idea of the proposed method and from our assumptions from related work we can derive main features which have to support:

1. Parallel collaborative work on software models in modules or layers \cite{bib11, bib4, bib65, bib20} and similar to draw.io \cite{bib22} or GenMyModel \cite{bib41} online in Virtual Reality, increasing productivity and productivity in work compared to standard or offline and uninformed noncollaborative work, visualizing trace of the particular sequence of work of co-workers to know who, where and what they are working on (Figure~\ref{fig:history_tracking}). The layer could be defined as a set of diagram elements with particular commonality and collective operation \cite{bib65}. A layered diagram is a partially ordered set of sequential, alternative, or orthogonal layers \cite{bib65} and we could use a special object algebra of addition and subtraction \cite{bib62, bib64}.

2. Visualizing parallel whiteboards on which the other developers are working (collaborators can work on the same layer, too) supporting reduction of the complexity of large models with decomposing the extensive models to the real modules or components with particular structures, supporting other use cases, visualizing similar parts on the other component layers to not reinvent the wheel and reduce redundant and vague elements.
Whiteboards provide space and comfort to developers and prepare the environment with the partitioning of the model into several worlds or interconnected layers containing: particular modules, author and time versions, particular type clusters (e.g. GUI, Business services, DB services). Developers can also highlight patterns or antipatterns in separate layers or converge to a lean architecture using the DCI approach with abstract domain classes, roles and scenarios \cite{bib24} in separate tiers.

3. Visualizing tiers with model and source code behind the diagram in various object-oriented languages (python, Java, C++, ...) to create a structural source skeleton of the project.

\begin{table}[h]
    \centering
    \includegraphics[width=0.46\textwidth]{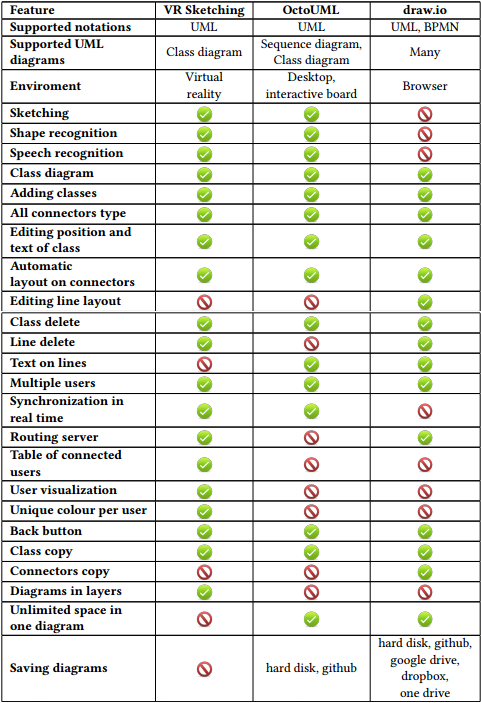}
    \caption{Comparison of our prototype with similar systems}
    \label{tab:coparison_of_tools}
\end{table}

\begin{figure}[h]
    \centering
    \includegraphics[width=0.48\textwidth]{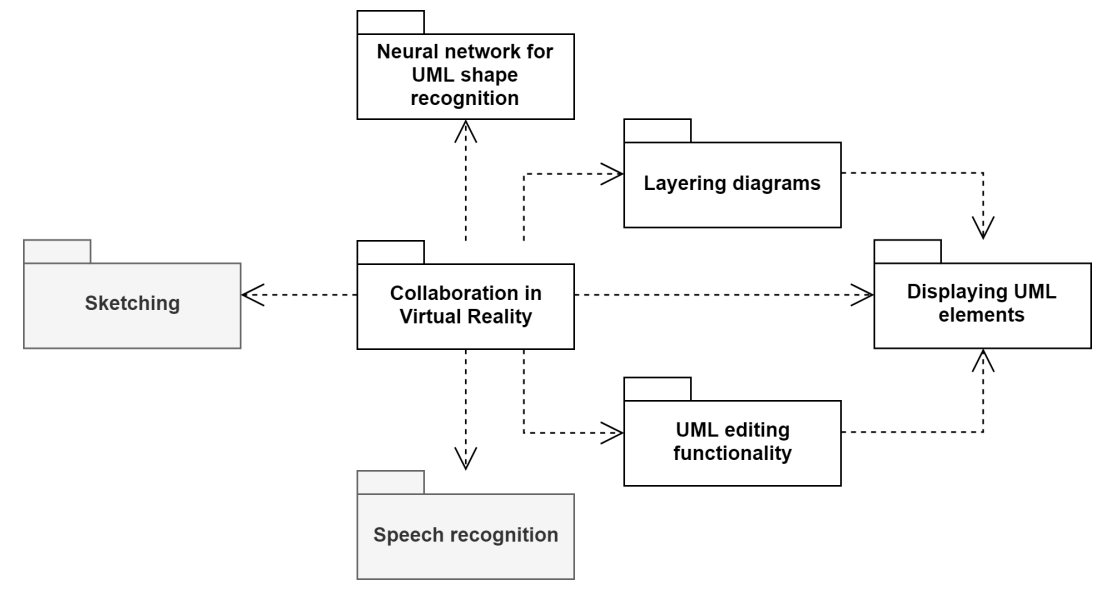}
    \caption{Package diagram of our prototype}
    \label{fig:package_diagram}
\end{figure}

To reduce the complexity of our prototype \emph{VRsketch}, we have decided to reuse open source modules for neural network and speech recognition (gray packages in Figure~\ref{fig:package_diagram}, white packages are implemented as a part of this project). 

Basic cognitive processes, such as focus and feelings (positive and negative), were indirectly investigated in post-task questionnaires using NASA TLX. Information processing, learning, and memory were investigated with RH2 and did not perform worse than the standard desktop, but neither did they perform better.

According to \cite{C1}, we can derive the most important cognitive processes in collaborative modeling from their detailed classification. These processes include solution search, problem and solution design, problem analysis, synthesis, comparing, judging, reasoning, evaluating, and decision making.

Comparing creativity and decision making in VR versus standard desktop systems was not our goal, but it is certainly an appropriate topic for future work where we would engage in a broader perspective with colleagues from psychology and cognitive science.  We were not motivated to increase creativity and decision making and did not pursue this. Instead, we have created a functional modelling tool in VR and tried to evaluate it against a similar desktop system. We did not want to end up like Yigitbas et al. \cite{bib37}, which was confirmed in the first phase of our research but after improving the user interface we got better results. Most importantly a good feeling prevailed in both phases of the experiments to motivate collaboration and satisfaction.

\subsection{Main functionality of our prototype} \label{sec3_2}

The prototype provides a useful tool for experimentation and evaluation of software modelling in VR. It includes the standard functionality of a collaborative software modelling tool:
\begin{enumerate}
\item Diagram editor (currently UML class diagram as the most used);
\item Support for collaborative parallel networking for multiple developers, Parallel collaborative work on software models in modules or layers;
\item Viewing collaborators as avatars, their movement at the whiteboard, view, hands and activity status (looking, drawing, erasing, pointing, speaking, etc.);
\item Displaying the work history of individual model designers in a unique color for each, which naturally fades over time;
\item The ability to display multiple tables and rearrange them;
\item Saving the results and uploading them to the system;
\item Ability to edit the diagram using VR keyboard and voice to insert text in addition to, Updating Data in Class Diagram Using Speech Recognition;
\item Image Recognition, Automatic recognition of sketches as diagram elements for more convenient and faster creation in VR;
\item Ability to teleport in VR space to individual charts;
\item Possibility of communication, voice-call over tool;
\item Visualizing parallel whiteboards on which the other developers are working;
\item Possibility of a deep and shallow copy of classes or entire collections and bundles of classes.
\end{enumerate}
All these features have been tested and tuned to standard to make it possible to work in VR comfortably and as fast as possible.

\subsection{Sketching and Editing with Tracking of Work History} \label{sec3_3}

We currently offer our prototype with the possibility to model in a UML Class Diagram. For recognition of UML shapes, we used and trained artificial neural network model from the machine learning (ML) library TensorFlow  \cite{bib14}. The external web service receives points and sends the possibility of each shape separately. Due to the delay in calling the web service and recognition, it is necessary to call it with a new thread and not block the smoothness of the application logic. We optimize the points by trying to send them within almost the same distance. If the user is drawing a line too fast, we split the line into more points. However, the hand of the user is not stable and shakes slightly while drawing. If the user is drawing too slow, virtual reality tracks the controller very specifically creating point clusters difficult to distinguish by neural network. We solved this problem by defining the minimum distance of the next collected point. 

Due to the ML tool's dependence on network resources and availability, we decided to implement a simplified and faster approach for shape recognition in the new version. This decision aimed to reduce noise in the experiment, as the initial prototype sometimes inaccurately evaluated shapes, or the ML service experienced unavailability issues.The new algorithm is specifically adapted to the class diagram and expects only lines and squares as input. It considers any drawing that starts and ends in the class elements as a line, while accepting all closed paths (start and end in the same place) as class shapes. All other drawings are not evaluated and are treated as informal sketches or drawings.

\begin{figure}[t]
    \centering
    \includegraphics[width=0.48\textwidth]{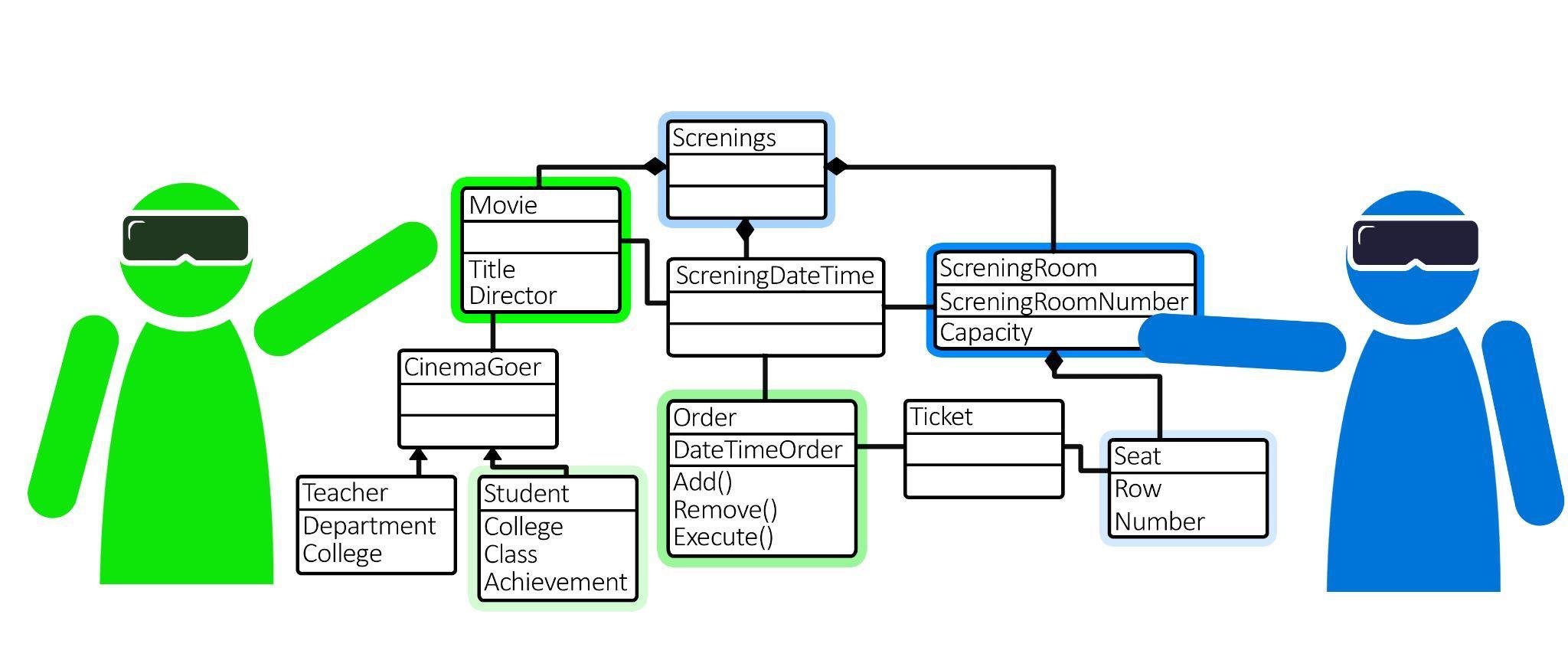}
    \includegraphics[width=0.2\textwidth]{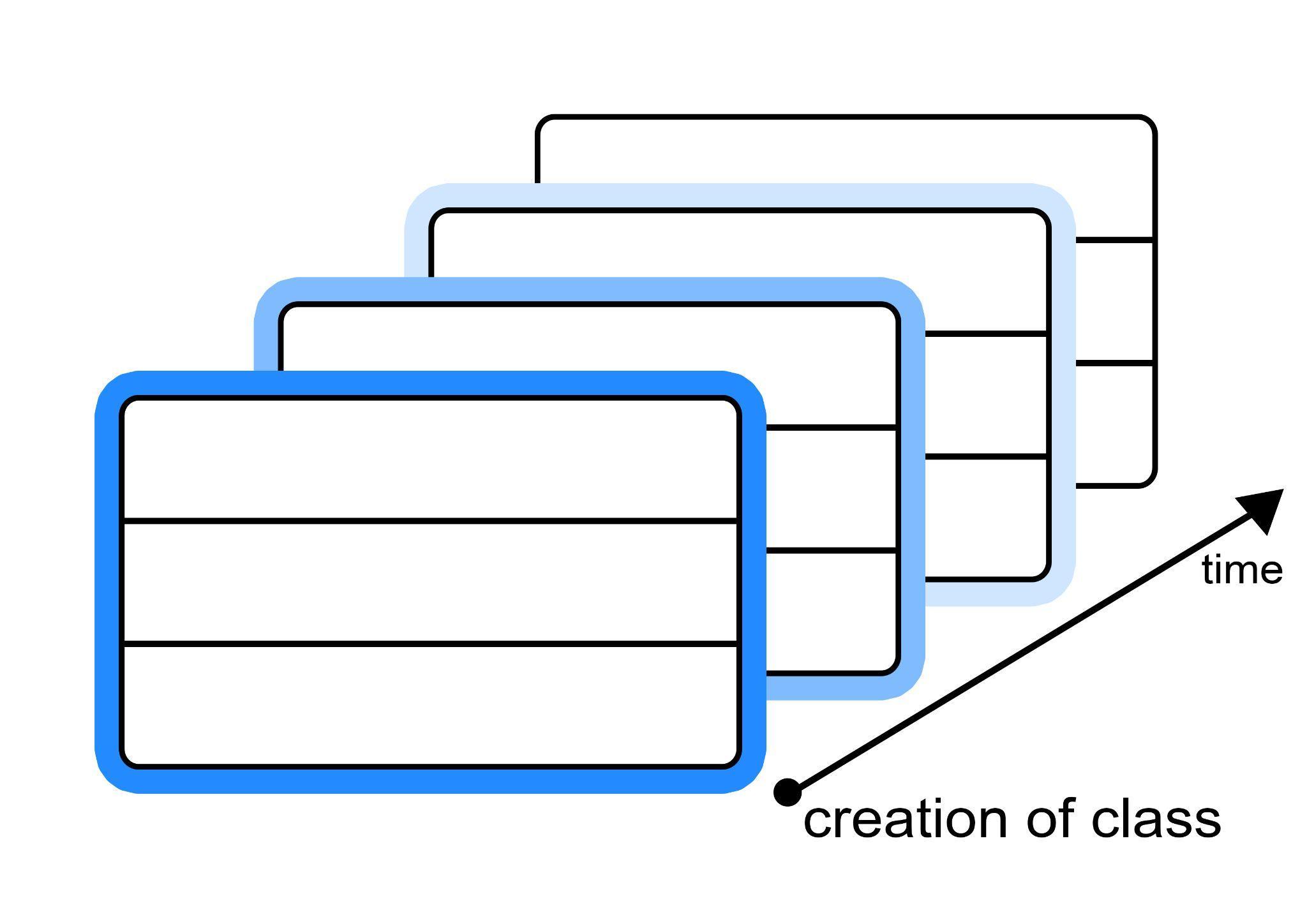}
    \caption{Tracking of the work history of blue and green developer}
    \label{fig:history_tracking}
\end{figure}

Each user is identified by a unique color in the space of virtual reality and diagram. This color is used every time, when a user does some change on the whiteboard. After the change, there is a fade function, which makes the borders fade until they lose color and stay black (Figure~\ref{fig:history_tracking}) and thus leaves a colored trace after each user from which we can sense and recognize his intention and way of thinking and the process of creating a model.

We decided to use the open source project, Drawing-VR, which supports painting in virtual reality. However, it was necessary to modify many of its features. We changed the painting method to collecting points, which are easy to synchronize over the network to particular users.

\begin{figure*}[t]
    \centering
    \includegraphics[width=0.266\textwidth]{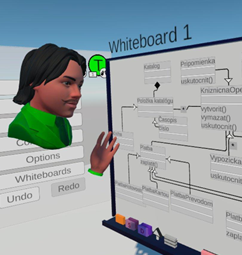}
    \includegraphics[width=0.309\textwidth]{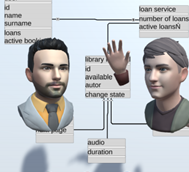}
    \includegraphics[width=0.332\textwidth]{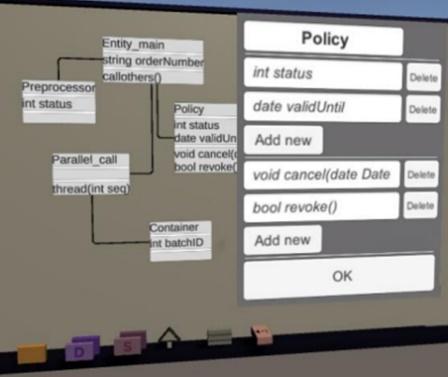}
    \includegraphics[width=0.437\textwidth]{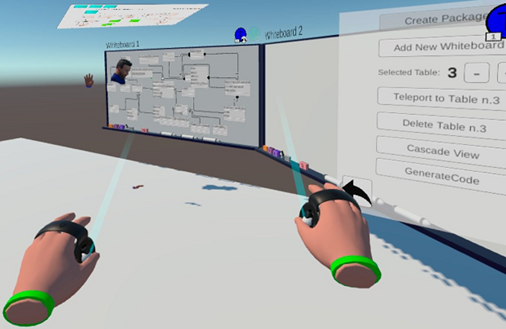}
    \includegraphics[width=0.418\textwidth]{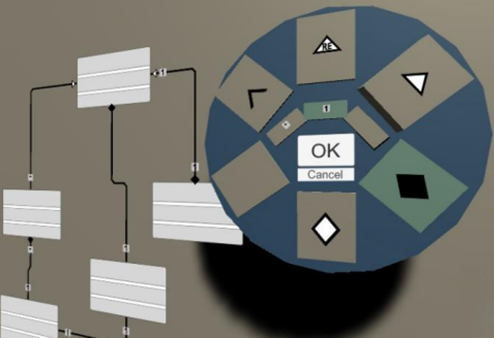}
    \caption{New version of avatars in VRsketch (a,b), GUI for editing class properties (c), improved settings of whiteboards and teleports (d), new association setting (e)}
    \label{fig:new_vrsketch}
\end{figure*}

\subsection{Improvements of the prototype after the first phase of experiments}\label{sec3_4}

\begin{figure}[t]
    \centering
    \includegraphics[width=0.48\textwidth]{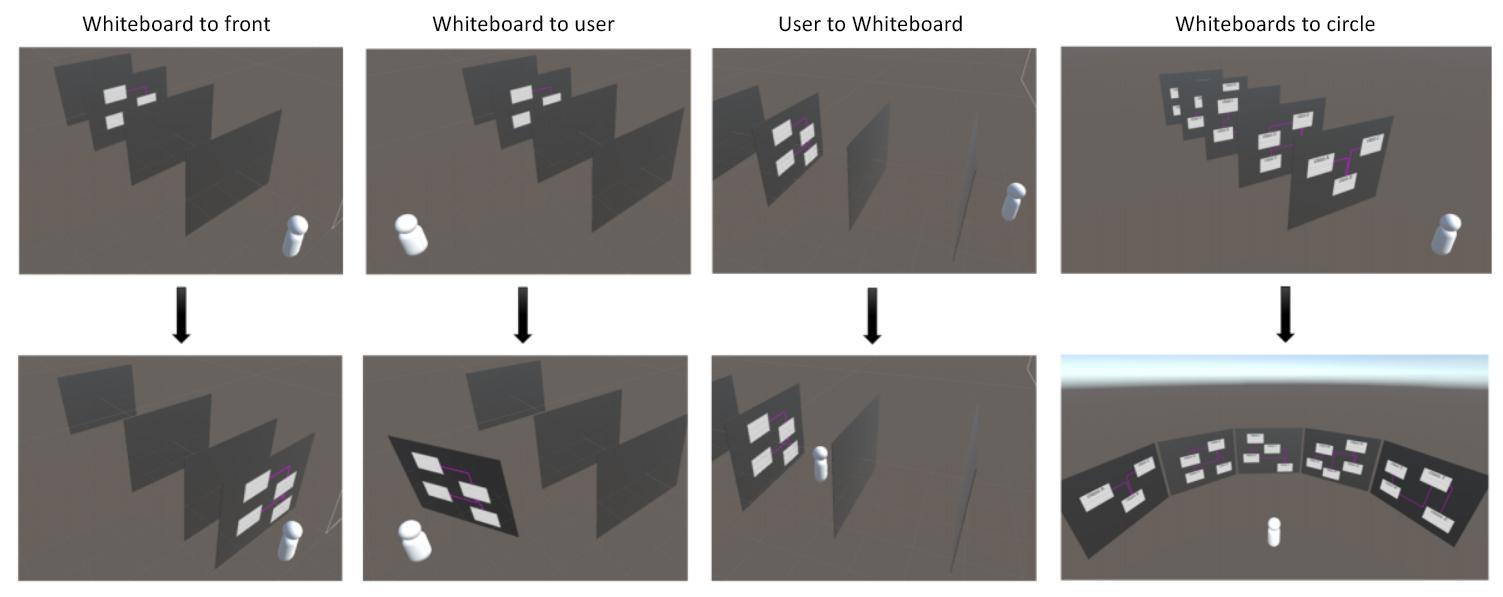}
    \caption{More opportunities for whiteboards and users to move}
    \label{fig:moving_boards}
\end{figure}

We were able to understand the importance of evaluation in this area \cite{bib71} and the necessity of improving the proposed visualization method, its prototype and interface \cite{bib69} from the surprising results we have obtained in the first phase of evaluation: rejection of the first and second hypotheses forced us to improve our prototype, enrich it with new features, and optimize the interface. We made the following list of new and redesigned features:

\begin{enumerate}
    \item new approach for sketching, creating and removing classes,
    \item added new and more detailed avatars (Figure~\ref{fig:new_vrsketch} a, b),
    \item changed interaction with the whiteboard (by deleting the panel and adding a menu into the table)
    \item changed the option for choosing relationship with UI panel for selecting new symbol and cardinality (Figure~\ref{fig:new_vrsketch} e),
    \item upgraded speech recognition (using more precise version from Google Cloud Services),
    \item changed the GUI of editing class properties (header, attributes and methods) with a new, straightforward window  (Figure~\ref{fig:new_vrsketch} c),
    \item added the possibility of a deep \& shallow copy (we ensured that the user can easily copy classes and set references),
    \item added the possibility of creating packages for reducing complexity,
    \item improved whiteboards settings, teleports  (Figure~\ref{fig:moving_boards} and Figure~\ref{fig:new_vrsketch} d), and upgraded voice-call over tool.
\end{enumerate}

\section{Experiment and evaluation}\label{sec4}

\subsection{Evaluation Concept}\label{sec4_1}

The main goals of our evaluation are to validate our approach and its tool and to compare it with other standard desktop tool collaborations‘ efficiency, immersion, and satisfaction. Our research hypotheses will also be derived from this intention.

\subsection{Experiment design}\label{sec4_2}

For the evaluation of collaborative modeling in Virtual Reality with our tool VRsketch we used a controlled experiment. In this experiment, the developers try to complete a software model with VRsketch and with a similar desktop application for collaborative modeling OctoUML v.4.2. Our goal is to compare satisfaction, comfort, cognitive load, effectiveness, and functionality of our prototype to a similar solution existing in a standard desktop setup.

Our controlled experiment involves two versions of the interaction technique (standard desktop application versus VR) with different user interfaces (OctoUML v.4.2 versus VRsketch v.2 and VRsketch v.3 in the second phase of experiment) as the \textbf{manipulated independent variable} \cite{BB}. The \textbf{response dependent variables} include task completion time, the number of syntactic errors, and the answers in the questionnaires (testing memory and satisfaction with collaboration), which are observable, easily measurable, and quantifiable using post-task questionnaires \cite{BB}.

We expose participants to two interfaces, VRsketch and OctoUML, and task them with solving prepared assignments. Subsequently, we compare the results obtained from each interface. We aim to reduce \textbf{confounding variables} by fixing the environment and computer devices for the experiment as \textbf{control variables}, maintaining a nominal setting \cite{BB}. The attributes of our experiment participants, such as gender, age, knowledge, profession, and VR experience, are considered \textbf{random variables}.

\begin{figure*}[b]
    \centering
    \includegraphics[width=0.9\textwidth]{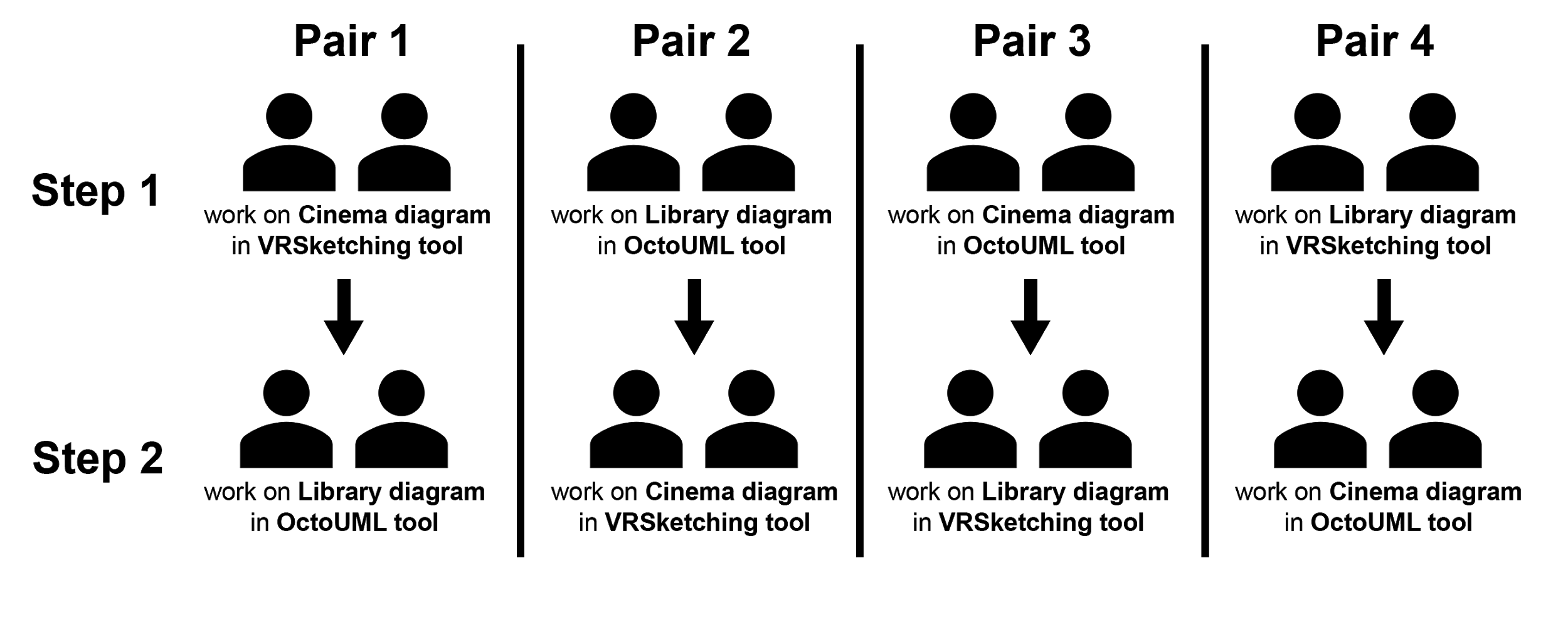}
    \caption{Schema of the whole evaluation process}
    \label{fig:schema_of_whole_experiment}
\end{figure*}

We strived to establish equal working conditions for both groups, similar tasks and equal preparation time. We created 2x2 pairs (8 people) that covered all combinations of tasks and sequences of tools (Figure~\ref{fig:schema_of_whole_experiment}). Therefore, we tried to get as many eights as possible, created and divided randomly from diverse backgrounds and varying levels of VR experience, including researchers, students, developers from the software industry, and professionals from small and medium software houses (see Table~\ref{tab:participants_structure}).

We chose OctoUML as a classic desktop modeling tool for its ease of use and efficiency. We used it for comparison with our VRsketch in our experiments because we assume that the other classic desktop tools such as DrawIO or Lucid Chart would have similar average times of task execution, but OctoUML uses sketchy diagrams with transformation into formal UML notation as VRsketch does and OctoUML supports interactive whiteboards. At the beginning of the experiments in 2019/2020, we did not know VmodlR \cite{bib37}, which is a similar tool in VR.

Currently, we evaluated our prototype with SUS \cite{bib36}, NASA TLX \cite{bib16} and as a comparison with similar applications (see Table~\ref{tab:coparison_of_tools}). For the experiment we selected OctoUML\footnote{https://github.com/Imarcus/OctoUML} \cite{bib1} that is a research tool for collaboration and sketching in standard environments.

NasaTLX \cite{bib16} is a method of evaluating subjective factors (overall workload, task difficulty, time pressure, exhaustion, etc.) on a user who performed some observed activity. This questionnaire focuses on mental, physical and temporal demand, performance, effort, frustration, and stress level.

\subsection{Research Hypotheses}\label{sec4_3}

Based on our assumptions in Chapter 3 that developers might work faster, be more immersed, and have a better feeling of collaboration in VR, we propose the following three research hypotheses RH1, RH2, and RH3.

\begin{description}\itemsep0em
    \item[\(\mathbf{RH1_0}\)] Collaborative Sketching in \textbf{VR is not slower} compared to standard tools (e.g. OctoUML) 
\end{description}

We will compare execution time of fulfilling a task in both tools VRsketch and OctoUML (i.e., duration of experiment). 

\begin{description}\itemsep0em
    \item[ \(\mathbf{RH2_0}\)] after collaborative modeling in VR, \textbf{developers will remember more design information} from the whole model compared to standard tools (e.g. OctoUML)
\end{description}

We will compare the results of the post-task questionnaires in both VRsketch and OctoUML tools.

\begin{description}\itemsep0em
    \item[\(\mathbf{RH3_0}\)] developers \textbf{prefer} collaboration in the VR environment
\end{description}

We will compare results from post-task questionnaires in both the VRsketch and OctoUML tools.

\subsection{Apparatus}\label{sec4_4}

For the experiment, we used regular standard computers with larger screens. Standard headsets with microphones were used for communication.

Windows and Linux were used as operating systems but they played no role as the participants did not come into contact with its features. 

The only relevant software was OctoUML for desktop computers and VRsketch for virtual reality.

In the first phase of the experiment (v1) we used two HTC Vive devices for VR, then in preparation step (v2) we used HTC Vive and Oculus Quest 1, and in the final phase we used 2x Oculus/Meta Quest 2.

\subsection{Tasks and Models for Experiment}\label{secC4_5}

We need to compare two collaborative tools (VRsketch, OctoUML) with two assignments in our evaluation process. Therefore, we need 2x2 couples in two steps for the following all combinations of couples (1, 2, 3, and 4), tools (X, Y) and assignments (A, B). For these four couples, we need eight developers. 
\begin{itemize}
    \item the first step: 1XA, 2XB, 3YA, 4YB,
    \item the second step: 4XA, 3XB, 2YA, 1YB.
\end{itemize}

\begin{figure}[h]
    \centering
    \includegraphics[width=0.47\textwidth]{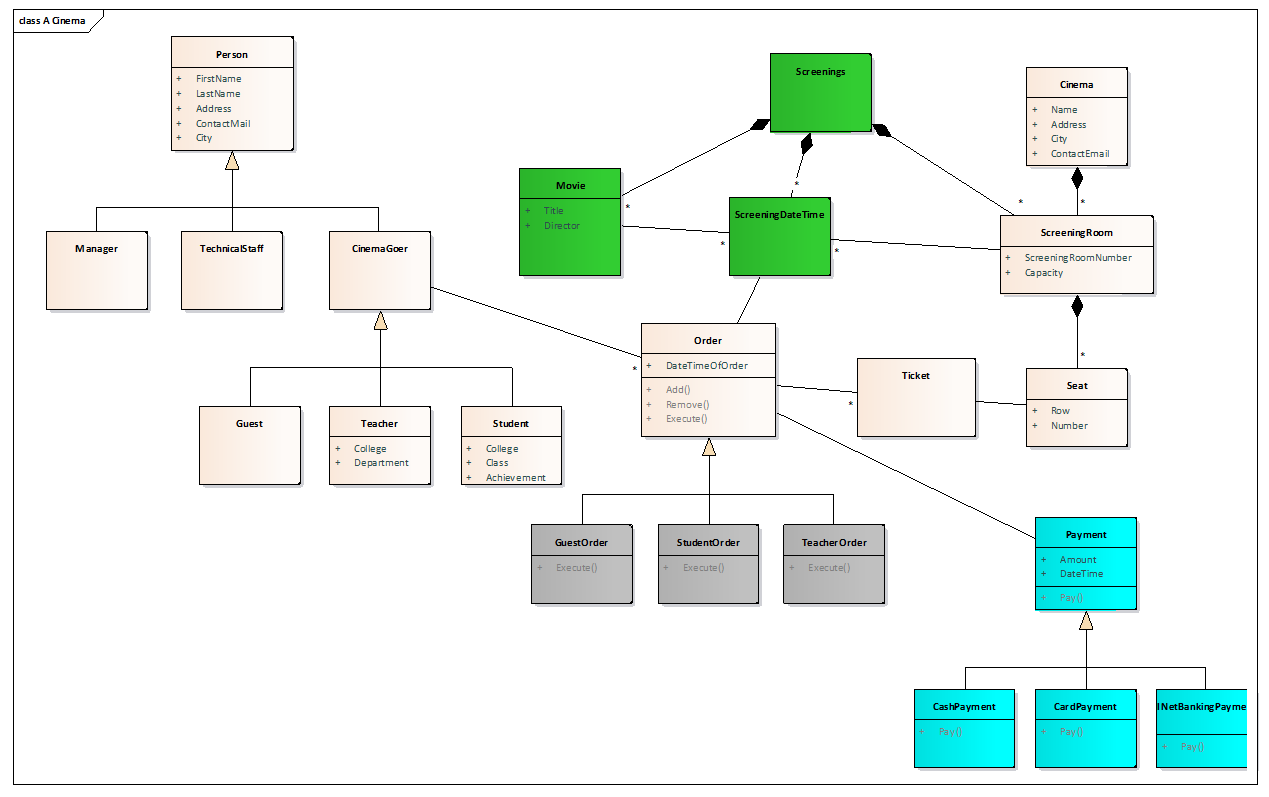}
    \caption{Class diagram \textit{Cinema} describing the basic schema of a cinema club}
    \label{fig:cinema_assigment}
\end{figure}

\begin{figure}[b]
    \centering
    \includegraphics[width=0.47\textwidth]{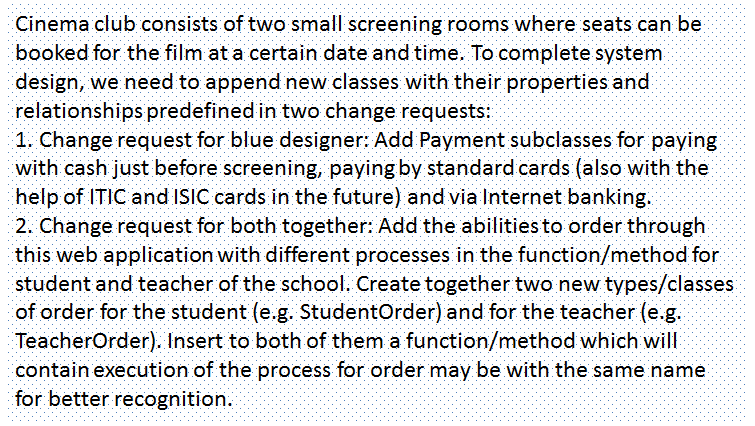}
    \caption{Assignment for blue designer}
    \label{fig:cinema_assigment_blue}
\end{figure}

\begin{figure}[t]
    \centering
    \includegraphics[width=0.48\textwidth]{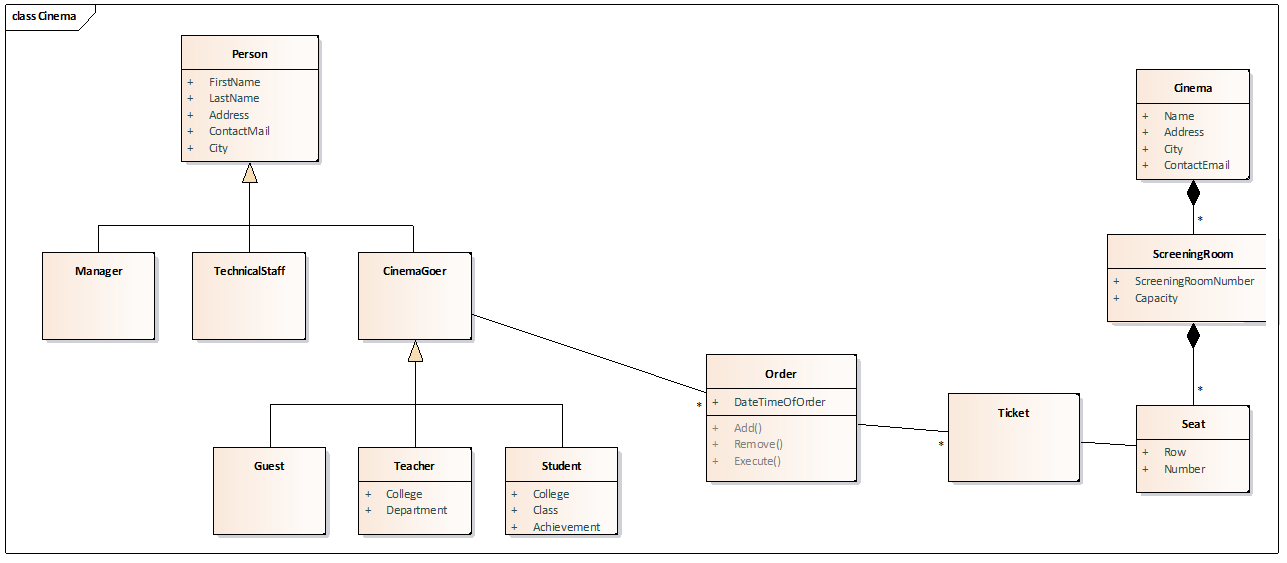}
    \caption{Class diagram \textit{Cinema} without green, blue and grey classes as a part of evaluation process}
    \label{fig:cinema_diagram_empty}
\end{figure}

We created two models for two assignments: \textit{Cinema} and \textit{Library}. In both assignments, the pairs consist of \textit{blue designer} and \textit{green designer}. Diagram \textit{Cinema} in Figure~\ref{fig:cinema_assigment} describes the basic model of the cinema club.

Diagram in Figure~\ref{fig:cinema_diagram_empty}  that participants received before the task is without \textit{blue}, \textit{green} and \textit{grey} classes (without classes that have to be completed through experiment). \textit{Blue}, \textit{green} and gray parts have the same degree of difficulty: sum of classes, attributes, methods, and relationships. Assignment for blue designer is in Figure \ref{fig:cinema_assigment_blue}.

\subsection{Procedure}\label{sec4_6}

\begin{figure}[!b]
    \centering
    \includegraphics[width=0.42\textwidth]{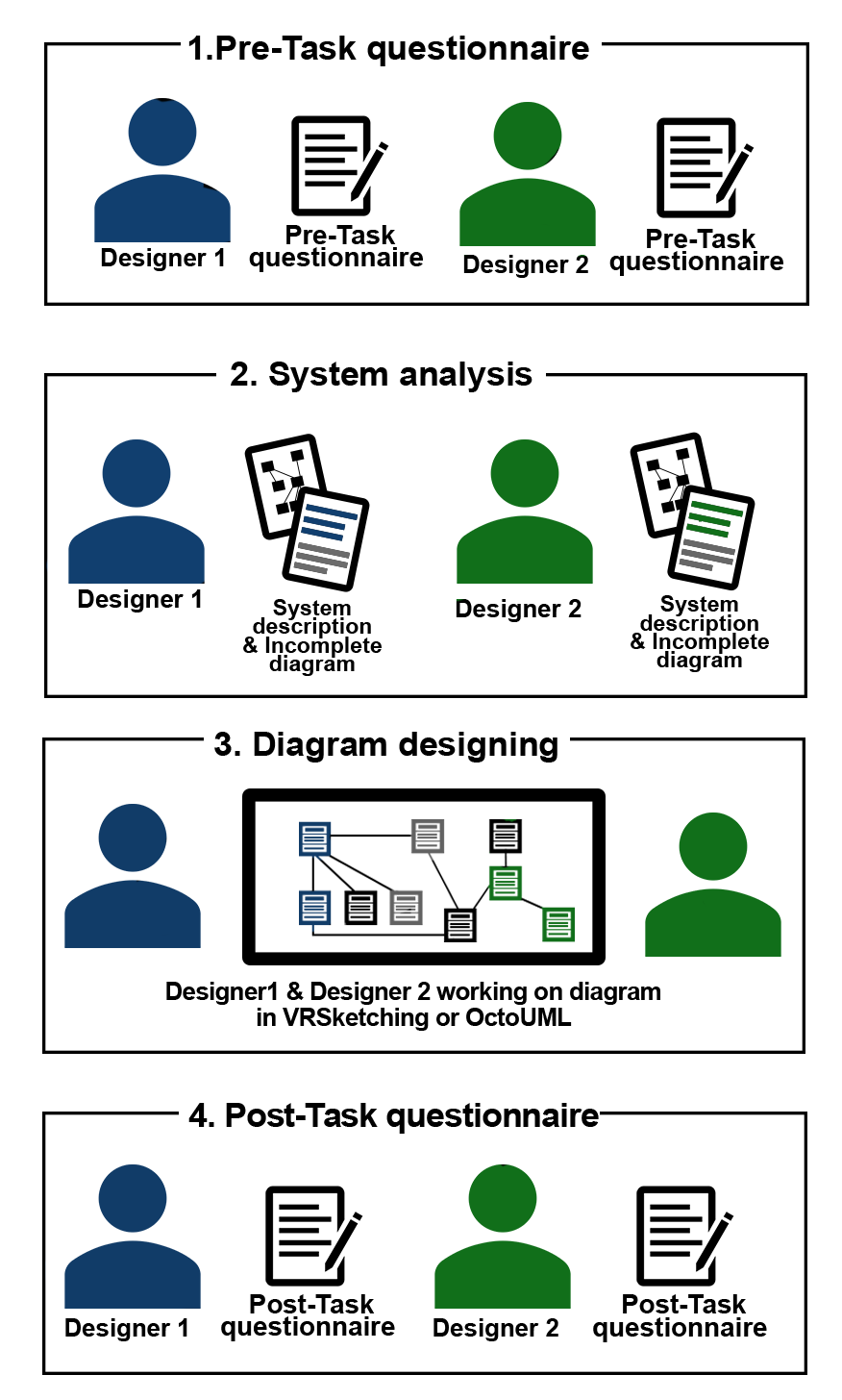}
    \caption{Schema of particular experiment in the evaluation process}
    \label{fig:schema_of_particular_experiment}
\end{figure}

The process of the experiment was divided into four steps (Figure~\ref{fig:schema_of_particular_experiment}), where the developers filled out pre-task questionnaires, analyzed the model and system, worked on the common model (main task), and completed post-task questionnaires.

\begin{figure*}[b]
    \centering
    \includegraphics[width=0.283\textwidth]{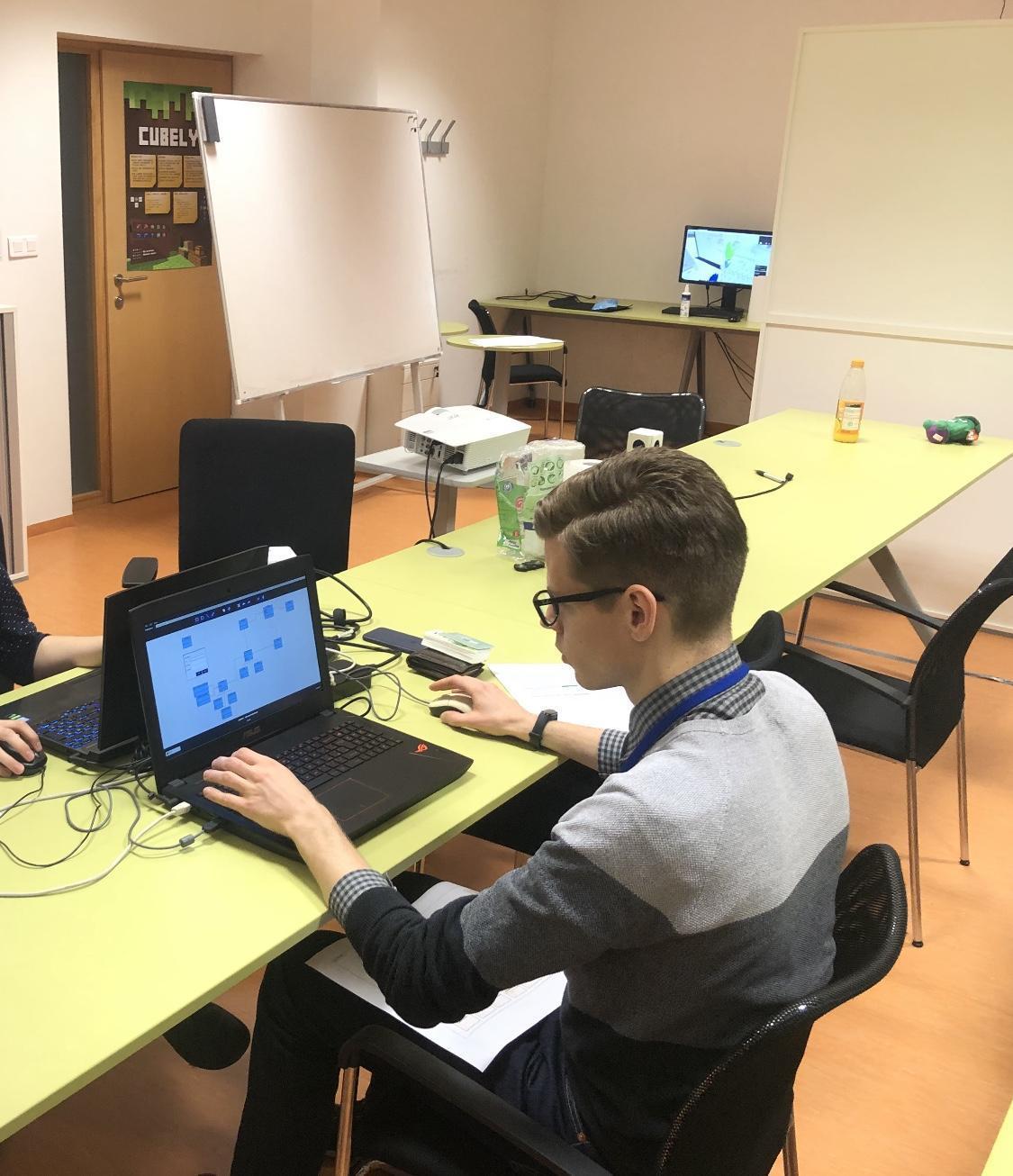}
    \includegraphics[width=0.45\textwidth]{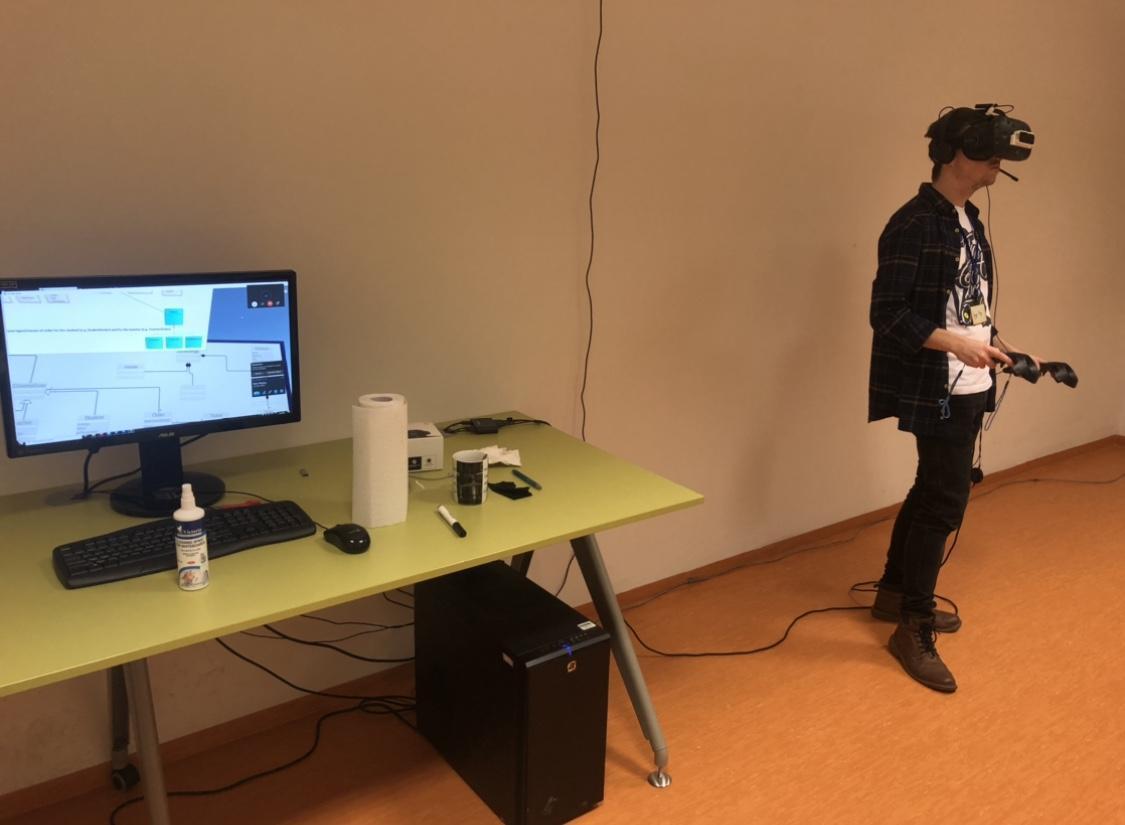}
    \includegraphics[width=0.218\textwidth]{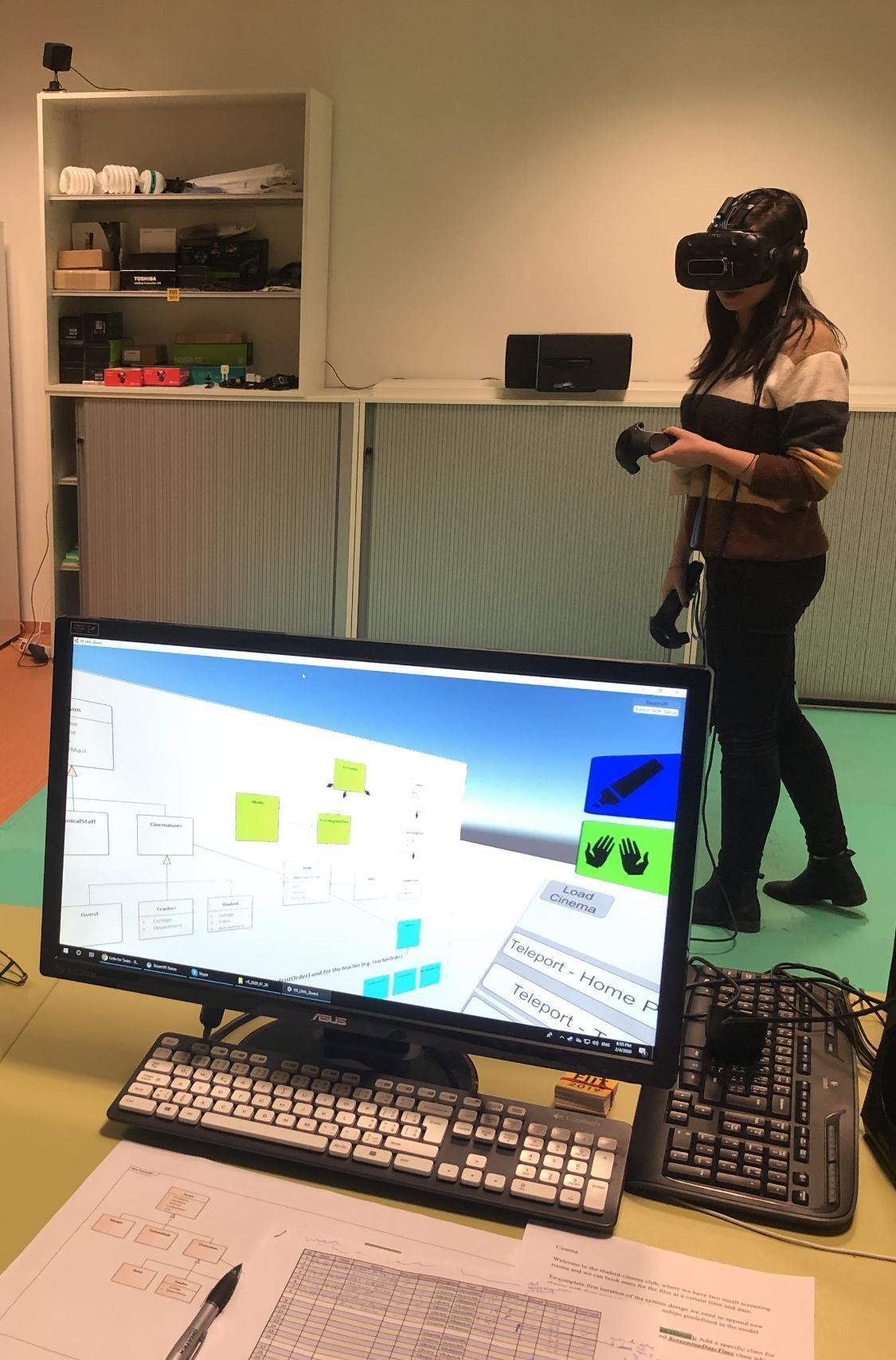}
    \caption{Stages for OctoUML (a) and VRsketch (b,c)}
    \label{fig:experiment_stages}
\end{figure*}

The main task was divided into two subtasks:
\begin{enumerate}
    \item Designer1 has to create a new part of the system (new classes with methods and relationships) and Designer2 has to create another part. In this subtask, we are observing their work on a separate problem (effectivity of tool editor, effort, etc.).
    \item Second subtask contains collaboration on a mutual problem - developers have to create another part of the system together (new classes with methods and relationships). We are observing their communication along with cooperation on a common assignment.
\end{enumerate}

\noindent Detailed description of the evaluation process:
\begin{enumerate}
    \item \textbf{Explaining purposes of the evaluation} to test the possibilities of Virtual Reality and not to test the knowledge and experience of the participants. We prepared a short presentation on the OctoUML and VRSketch functionality.
    \item \textbf{Creating random couples} to achieve more independent statistical results. In real life situations, senior designers and newcomers do not know one another either, so they are not well coordinated. In January 2020, we created 21 random pairs from 42 participants, and each pair worked with both tools (VRSketching and OctoUML) on different models (Figure~\ref{fig:schema_of_whole_experiment}). We have obtained 84 results for comparison (42 from the VRsketch experiment and 42 from the OctoUML experiment). Participants used the names of their favorite actresses, actors, and singers from our list as a unique and unrepeatable token to anonymize them not only for GDPR purposes, but also to feel comfortable and stress free.
    \item \textbf{Preparing tools and equipment}. After introduction we divided two pairs into separate spaces or rooms and prepared the necessary tools and equipment (two desktops for OctoUML, four desktops for pre-task and post-task questionnaires and assignments, two VR stages with headset) for them to achieve the best performance (Figure~\ref{fig:experiment_stages}).
    \item \textbf{Pre-task questionnaires}. Couples had to complete a questionnaire focused on their knowledge and experience in UML, UML tools, OOAD, and English language level (see Figure~\ref{fig:fig_pre-questions}).
    \item \textbf{Tutorial}. Lack of experience with Virtual reality (especially with UML class diagram sketching in VR) and OctoUML tool has forced us to create a tutorial focused on handling and manipulation with both of these tools. Tutorials consisted of tasks, where the developers were introduced with functionality of tools (creating/deleting classes, editing methods or attributes, creating/editing/deleting relations, copying classes, speech recognition or sketching) and tried them. The tutorial was written in the form of 10-20 short commands.
    \item \textbf{Model and Assignment}. Participants received assignments and a model with missing parts which they had to complete and a short description of the model. They needed to analyse and understand the model, think about the missing classes, identify them and their position in the model.
    \item \textbf{Collaboration and modeling}. Participants are working on the first task in the first tool after proper analysis of model and assignment. In the case of VRSketch, participants are working in separate rooms/stages with their own VR device. If they are working on OctoUML, they have two PCs with \emph{Skype} installed for online voice communication.
    \item \textbf{Post task questionnaires}. After finishing the task (and recording their duration time), the participants complete three post-task questionnaires. The first questionnaire (see Figure~\ref{fig:fig_post-questions}) is focused on their knowledge of the model they were working on (relations between classes, method/attributes in classes, inheritance, and whole structure of diagram). The second questionnaire (see Figure~\ref{fig:fig_post-questions3}) is focused on immersion, collaboration intensity, comfort, level of interaction, and general feeling about the tool and the modeling in it. The third questionnaire is the SUS (see Figure~\ref{fig:fig_post-questions2}) and NASA TLX questionnaire on mental, physical, temporal demand, performance, effort, and level of frustration. Post-tasks  are described in Chapter \ref{secC4_9}.
    \item \textbf{Swapping tools and assignments}. Steps 5,6,7,8 are repeated, but with different tools and models to create the same sets of tool and task combinations: (Octo:Task1 \& VR:Task2, VR:Task1 \& Octo:Taks2, Octo:Task2 \& VR:Task1, VR:Task2 \& Octo:Task1).
\end{enumerate}

\subsection{Participants in the first phase of experiments}\label{secC4_7}

\begin{figure}[h]
    \centering
    \includegraphics[width=0.48\textwidth]{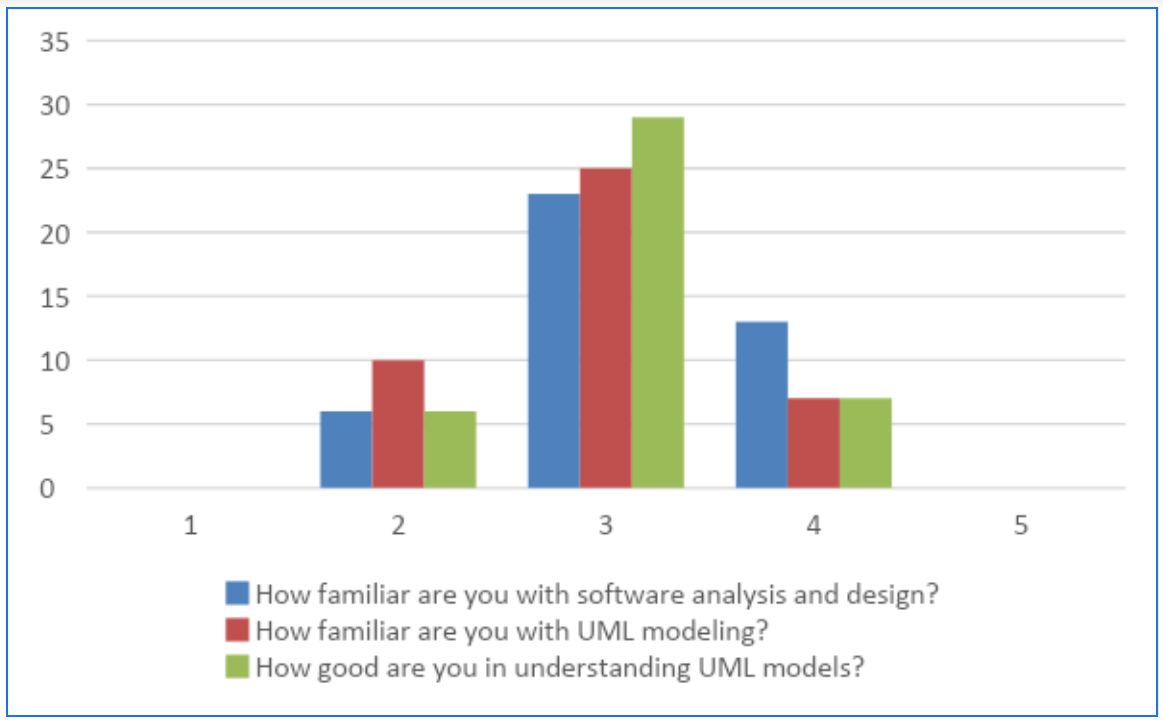}
    \caption{Responses of the participants to questions}
    \label{fig:particimant_answers}
\end{figure}

In the first phase we had 42 MSc students for 21 couples (more than 5 x 8 developers) and we achieved \textit{84 outcomes} (2 steps x 42 participants) from \textit{42 experiments} (2 steps x 21 couples). They were master's students from the Department of Computer Science at the Faculty of Informatics and Information Technologies (University of Technology). These students already had completed courses focused on analysis and design of software systems with UML notation. They were capable of creating or completing existing models because of their knowledge and circa three years experience. They represented newcomers and junior developers in the projects and software companies. The 42 participants were between 21 and 24 years old (12 participants female and 30 male). The graph in Figure~\ref{fig:particimant_answers} presents the answers to the selected questions on the Likert scale 1-5. We will try to improve the set of participants and add senior developers to the experiments in the final phase of the evaluation of our second version of the prototype.

\subsection{Participants in the second phase of experiments}\label{sec4_8}

In the second phase of experiments with the new version of the prototype, we had 80 participants and we achieved \textit{160 outcomes} (2 steps x 80 participants) from \textit{80 experiments} (2 steps x 40 couples). We used not only students from the departments of computer science at the Faculty of Mathematics, physics, and informatics, but also developers and senior developers from software companies. In addition to students, experiments were also supported by large software houses (Gratex, Softec, Ditec) and progressive small new startup companies (see Table~\ref{tab:participants_structure}). All 80 participants were between 21 and 54 years old. 50\% of them were programmers, the remaining 50\% were combined (analyst, tester, programmer, IT student, etc.). Altogether 17 participants were females and 63 were males.

\begin{table}[h]
\centering
\begin{tabular}{|l|c|}
\hline
\multicolumn{1}{|c|}{Source}     & \begin{tabular}[c]{@{}c@{}}Number of \\ Participants\end{tabular} \\ \hline
Gratex International, developers & 30                                                                \\ \hline
Computer Science Dep., students  & 30                                                                \\ \hline
Computer Science Dep., R\&D      & 4                                                                 \\ \hline
Kinit Institute, R\&D            & 4                                                                 \\ \hline
Me-Inspection, developers        & 2                                                                 \\ \hline
Ditec, developers                & 2                                                                 \\ \hline
Softec, developers               & 4                                                                 \\ \hline
Luigi's Box, R\&D                & 2                                                                 \\ \hline
Evolveum, developers             & 2                                                                 \\ \hline
\end{tabular}
\caption{Structure of participants}\label{tab:participants_structure}
\end{table}

\begin{figure*}[t]
    \centering
    \includegraphics[width=0.60\textwidth]{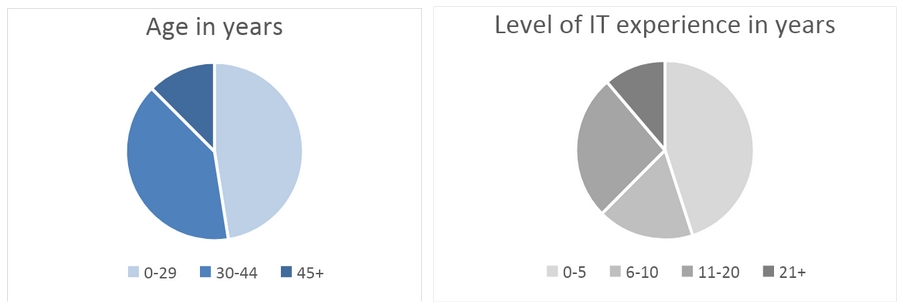}
    \includegraphics[width=0.60\textwidth]{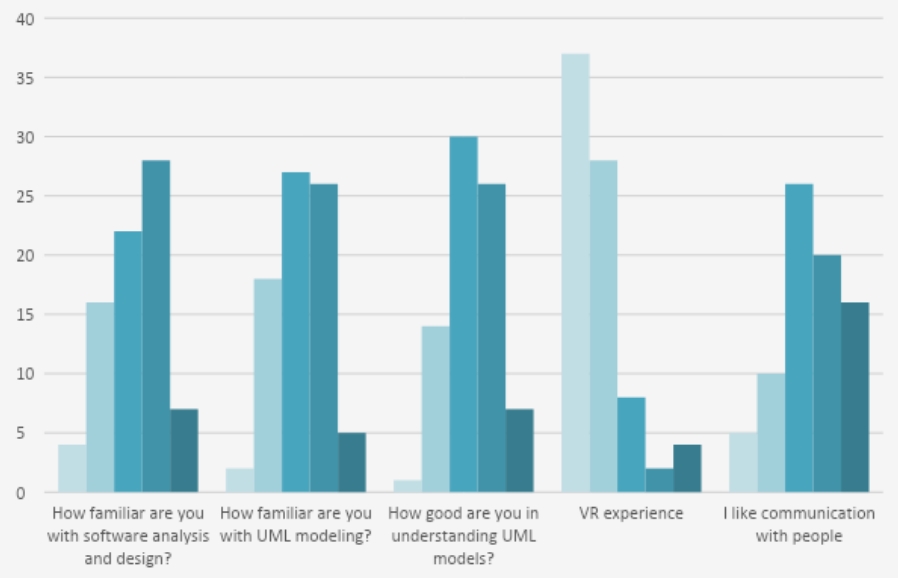}
    \caption{Demografic structure of participiants in age (a) and experience (b), and the answers to selected questions (c)}
    \label{fig:demografic_structure_of_participiants}
\end{figure*}

Graphs in Figure~\ref{fig:demografic_structure_of_participiants} present demografic structure of participiants in age and experience, and the answers to selected questions in likert scale 1..5.

With experience from the first phase of experiments, we can use the advantage of a larger number of participants and wider result set from the second stage of experiments to analyze correlations between age, experience, knowledge of UML or VR, etc. on one hand; and number of errors, satisfaction with collaboration in VR, quality of responses from the content of the model, on the other hand. Although in some cases (communicativeness, experience with UML or VR) the answers or satisfaction with VR were slightly better, we did not find a statistically significant correlation 
on which we could establish serious deductions and recommendations, according to the structure of the participants or developers in software houses.


Our repeated experiments took 2 months – several hours per day. After analysing the results and close observation of the participants’ activity it can be concluded that it is necessary to improve the training with/for VR tools to achieve better satisfaction and efficiency.

Controlled training lasts 15 minutes. It has to be extended in lege artis to several days creating a small real-world project for team members who want or have to work and communicate with others in VR tools. Of course, it is also necessary to put emphasis on improving communication when modelling in VR, but we could apply this to any social human activity. Well-developed communication skills from IT experts is a requirement which is not often met. However, there is hope that by using VR and a VR avatars, their approach to others might change, but this is not the scope of this paper.

\subsection{Pre and Post Questionnaires}\label{secC4_9}

Before any introduction or task, we were interested in their level of knowledge in software modeling, UML notation, understanding diagrams, and English language level. In this chapter we use Likert scale of answers from 1 (poor, not at all) to 5 (very good, very familiar). Every participant of the experiment was asked the following questions:
    
\begin{figure}[h]
    \centering
    \includegraphics[width=0.48\textwidth]{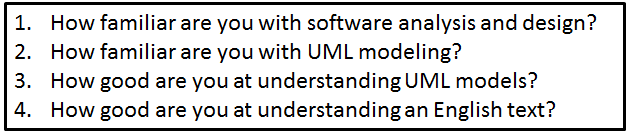}
    \caption{Sample of questions from pre-task questionnaires}
    \label{fig:fig_pre-questions}
\end{figure}

After completion of the task (working in VRSketching or OctoUML), participants were asked to fill out a post-task questionnaire divided into 3 parts.

\begin{figure}[h]
    \centering
    \includegraphics[width=0.48\textwidth]{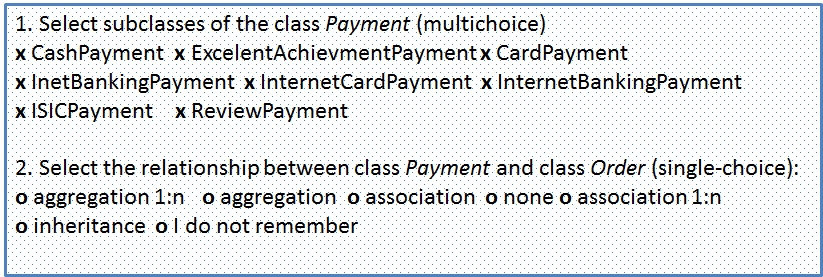}
    \includegraphics[width=0.48\textwidth]{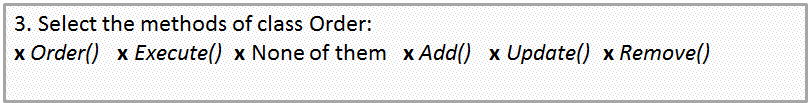}
    \includegraphics[width=0.48\textwidth]{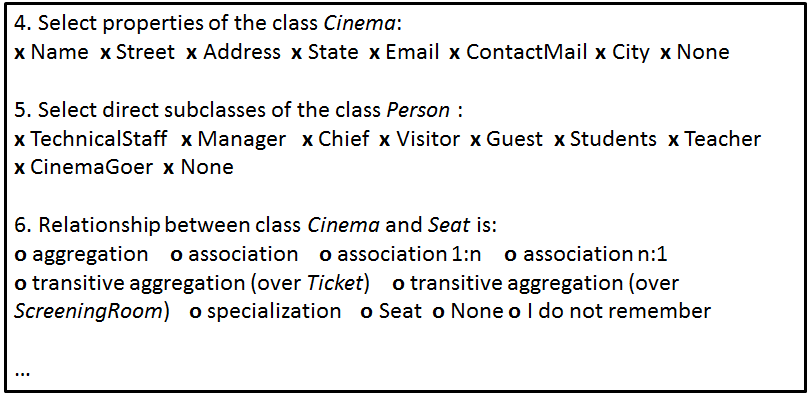}
    \caption{Sample of questions for RH2 testing from post-task questionnaires}
    \label{fig:fig_post-questions}
\end{figure}

In the first part, we need to analyze and compare their level of understanding of the models they had been working on (relations between classes, method and attributes in the classes) to find out the participant's immersion into the task in virtual reality and with the standard desktop collaborative modeling application. In this set of questions, the answers were single-choice or multiple choice (from 6 to 10 possible options). Every designer (green and blue) had his own set of questions for every particular model (Library and Cinema). For example, the questions for the blue designer in the Cinema model were as in Figure~\ref{fig:fig_post-questions}.

The questions are divided into the first blue part (new classes from the task of the blue developer), the second gray part (new classes from the common task with the green developer), and the last white part (already existing classes from the model).

\begin{figure}[b]
    \centering
    \includegraphics[width=0.48\textwidth]{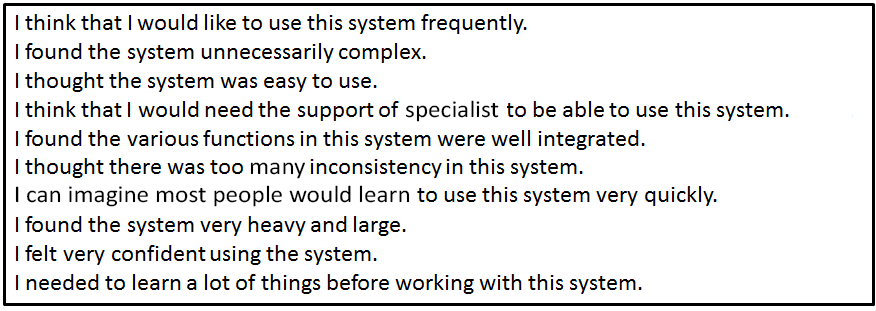}
    \caption{Sample of questions (SUS) from post-task questionnaires}
    \label{fig:fig_post-questions2}
\end{figure}

The second part of three post-task questionnaires was inspired by SUS \cite{bib36}. It was focused on the designer's opinion about the usage of a particular method and tool in real-life software development, about complexity, functionality, immersion, collaboration intensity, comfort, interaction level, overall feeling, and recommended improvements. A selection from these questions is shown in Figure~\ref{fig:fig_post-questions2}.

\begin{figure}[h]
    \centering
    \includegraphics[width=0.48\textwidth]{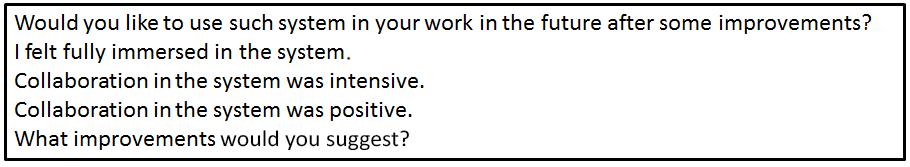}
    \caption{Sample of questions for RH3 testing from post-task questionnaires}
    \label{fig:fig_post-questions3}
\end{figure}

The last part of the post-task questionnaire (Figure~\ref{fig:fig_post-questions3}) was a NASA TLX questionnaire on mental, physical, temporal demand, performance, effort, and level of frustration \cite{bib16}.

\subsection{Results}\label{sec4_10}

In the first phase of experiments we obtained shorter completion time with the tasks in OctoUML (RH1 was not accepted). It seems, that newcomers in VR need more time to train with relatively new and unknown technology (interface in virtual reality). Only three pairs from the set of 21 couples performed better in VR than in OctoUML on standard PCs: they have (extensive) experience in VR from playing games in VR. 

We found that developers are more satisfied with collaboration in Virtual Reality and feel better than in a standard desktop environment. RH3 was accepted. This result was promising and we were motivated to improve our VR prototype and prepared a second phase of experiments.

The evaluation of the results from the new series of experiments with the second version of the VRsketch prototype is satisfying. Despite the Covid-19 pandemic and the lockdown with restrictions (home office and online teaching) we found more participants for the experiment with the second version of the prototype. 42 IT students took part in the first phase of the experiment and 80 students and developers in the second phase.

\begin{figure*}[t]
    \centering
    \includegraphics[width=0.99\textwidth]{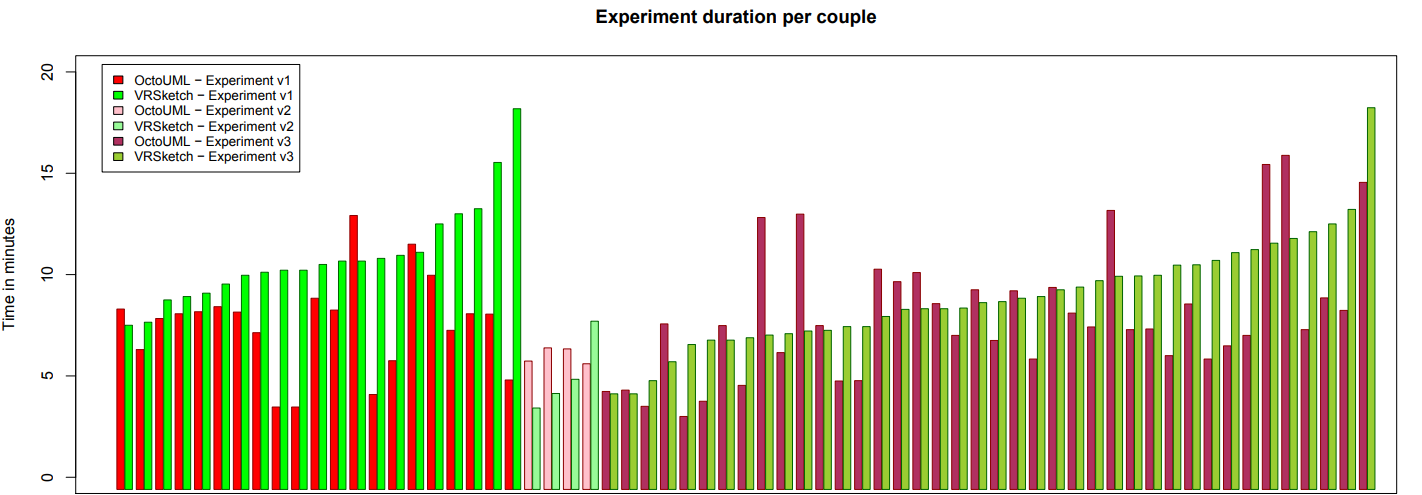}
    \caption{Duration of the experiment per couple (new results from the second evaluation on the right side)}
    \label{fig:experiment_duration_per_couple_2}
\end{figure*}

We were able to perform extensive experiments to confirm our expectations for higher efficiency (time to complete the tasks in VR) and satisfaction. The mean of time in VRsketch is 8:54 min. and the median is 8:39 min. The mean time in OctoUML is 8:01 and the median is 7:27. The spread with respect to time is slightly smaller for the standard application (min=3:00, max=15:53, diff=12:53) than for VR (min=4:07, max=18:14, diff=14:07). Figure~\ref{fig:experiment_duration_per_couple_2} shows 21 results from the first experiment (labeled v1) with 42 people, 4 results of the preparatory experiment (labeled v2) with 8 people and 40 results from the second final experiment (labeled v3) with 80 people.

We \textbf{can accept} $\mathbf{RH1_0}$ because the p-value is  - 0.4296 on the significance level $\alpha=0.05$ and the p-value $> \alpha$, therefore there is no statistically significant difference in favor of OctoUML between the times in the VRsketch and OctoUML.

The next Figure~\ref{fig:quality_answer_and_duration} shows a comparison of the results in the first and second experiment, where it can be seen that the quality of the answers did not change and it was approximately the same for the desktop OctoUML and the VR environment VRsketch (first two graphs on the left). The results of runtime duration (last two graphs on the right) were improved in the second experiment for VRsketch and they were already statistically insignificant although they were still a little worse for VR than for desktop.

\begin{figure*}[b]
    \centering
    \includegraphics[width=0.40\textwidth]{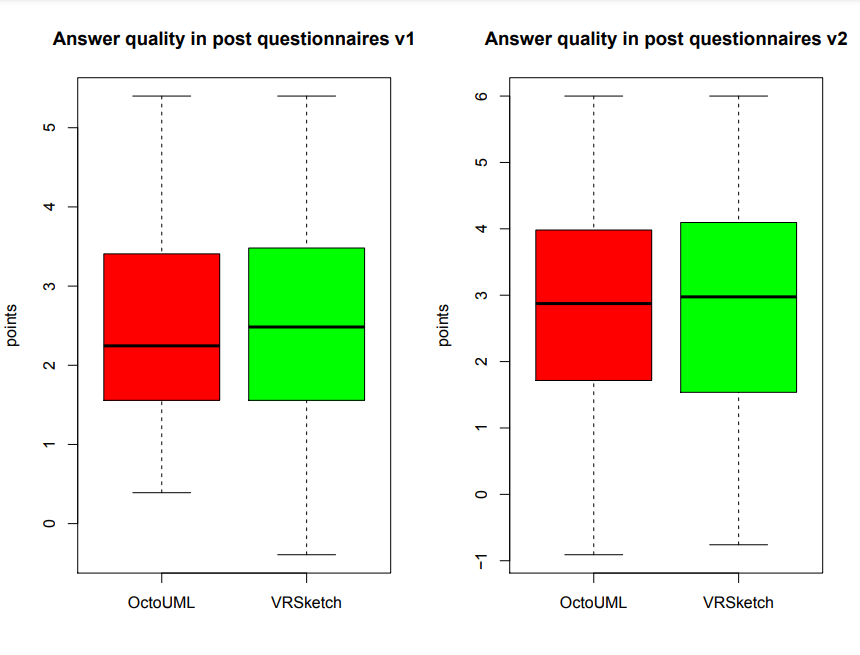}
    \includegraphics[width=0.40\textwidth]{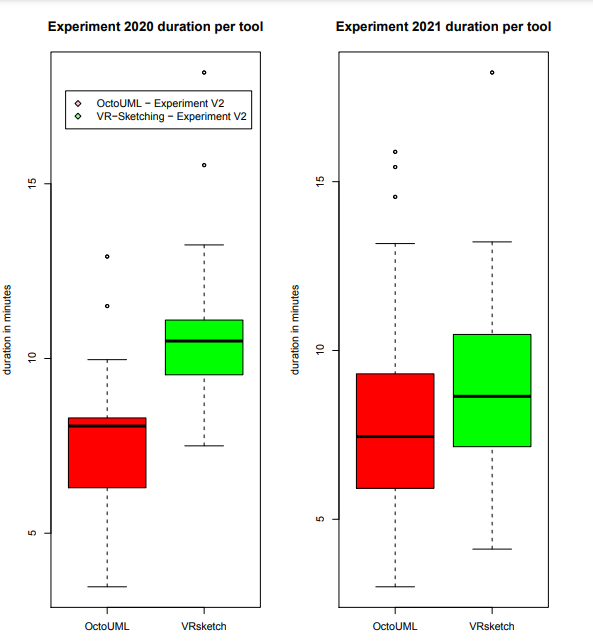}
    \caption{Results of the first and second phases of the experiments: quality of answers and duration per tool}
    \label{fig:quality_answer_and_duration}
\end{figure*}

We \textbf{cannot accept} $\mathbf{RH2_0}$: after collaborative modeling in VR, developers will remember more design information from the whole model/functionality in contrast to standard tools (e.g. OctoUML), because p-value is 0,741 on the significance level $\alpha=0.05$ and the p-value $> \alpha$, therefore there is statistically no significant difference between the answers in VRsketch and OctoUML in favor of OctoUML or VRsketch.

\begin{figure*}[b]
    \centering
    \includegraphics[width=0.90\textwidth]{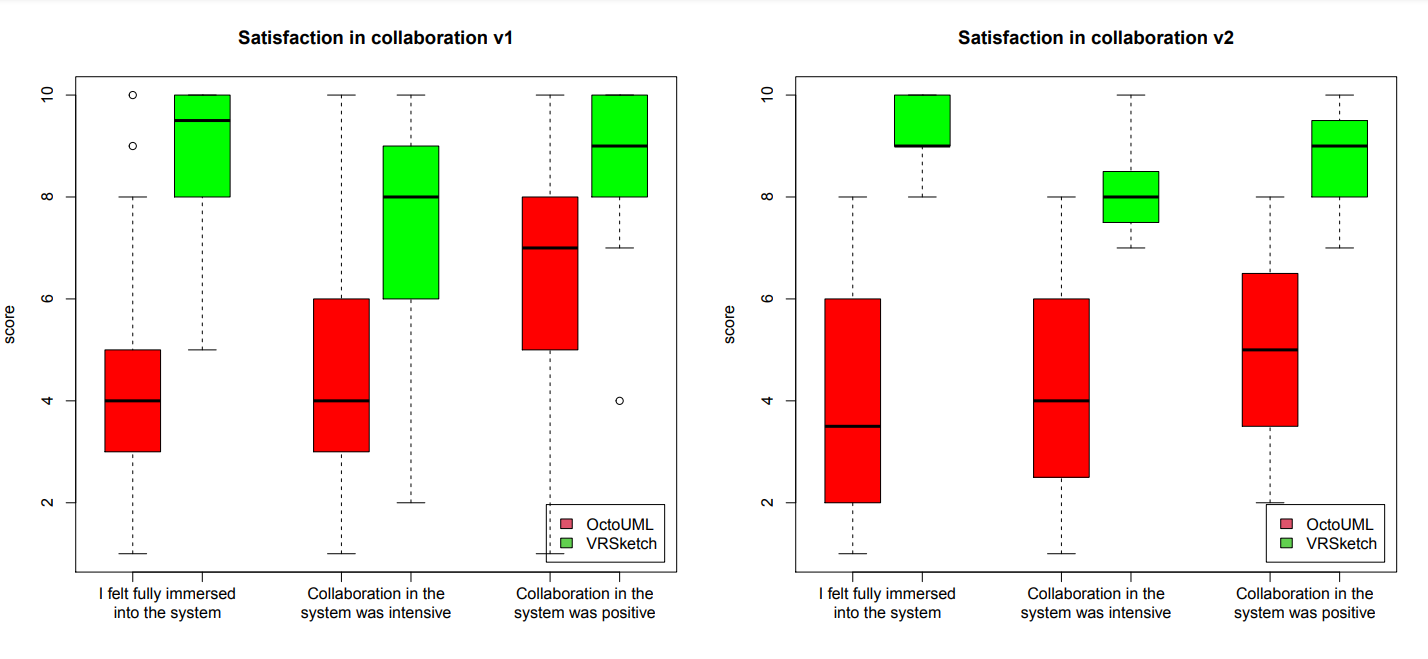}
    \caption{Aggregated results over the sum of specific questions from the first and the second phase of the experiments}
    \label{fig:agg_result_over_specific_questions}
\end{figure*}

In the third step we can compare immersion in collaboration, intensity and positive feeling of collaboration in VRsketch and in standard desktop tool (e.g. OctoUML). Both rankings contain 42 sums of seven answers in normalized scales (1 to 10). From the graph in Figure~\ref{fig:agg_result_over_specific_questions} it appears that developers are more satisfied with collaboration in VR and feel better than in standard desktop environments.

Specific questions in Figure~\ref{fig:agg_result_over_specific_questions} are the same collections as in the first phase of experiment, described in~\ref{secC4_9}. Figure shows that the satisfaction with VRsketch was still higher with VRsketch than with OctoUML in the second experiment. The difference between the values of satisfaction with VRsketch and OctoUML were larger.

We \textbf{can accept} $\mathbf{RH3_0}$: developers \textbf{prefer} collaboration in a VR environment, because p-value is 1.197e-06  on significance level $\alpha=0.05$ and p-value $< \alpha$, therefore there is statistically significant difference between rankings of VRsketch and OctoUML in favor of VRsketch.

\subsection{Resume}\label{sec4_11}

After the end of the second series of experiments in 2022, we analyzed the data obtained and found the following positive facts about the new version of VRsketch:

\begin{enumerate}[label=(\Alph*)]
    \item The execution times in \emph{VRsketch2} (2021/22) are better than the times in \emph{VRsketch1} (2020).
    \item Execution times in \emph{OctoUML} are not significantly better than the times in \emph{VRsketch2} (2021/22).
    \item Satisfaction with \emph{VRsketch2} is still high, developers are more satisfied with \emph{VRsketch2} (2021/22) than with \emph{OctoUML} (RH3 is confirmed), and satisfaction with \emph{OctoUML} has decreased in the second series of experiments.
        \item Ratio of task execution times in both experiments was better with our VR tool (\emph{VRsketch}) versus the standard collaborative modeling tool (we used \emph{OctoUML}) than in experiments with a comparable VR tool (\emph{VmodlR} \cite{bib37}) versus the standard collaborative modeling (\emph{Lucidchart}): $T_{VRsketch} / T_{OctoUML} < T_{VmodlR} / T_{Lucidchart}$ ($8:39 / 7:27  <  13:42 / 07:35$).
    \item We achieved better results in the number of errors when editing the diagram during the experiment (1.775 errors on average per task in \emph{VRsketch} versus 1.850 errors per task in \emph{OctoUML}). In \emph{VmodlR} experiments \cite{bib37} they observed a mean error rate of approximately 1.5 error per minute versus 0.66 error per minute in the web application.
    \item Participants prefer \emph{VRsketch} to \emph{OctoUML} in two questions: 1. \emph{I think I would like to use this system frequently} (3.1750 vs. 2.9250 from 5 points in Likert scale), and 2. \emph{Would you like to use such a system in your work in the future after some improvements?} (3.4375 vs. 3.4000).
\end{enumerate}

According to points D, E and F, our participants worked almost as fast in VR as in the compared standard desktop environment, they did not make more mistakes and did not refuse to use VRsketch in the future.

We used similar libraries and the Unity environment as in VmodlR \cite{bib37} and we also found (as they did) that the work in VR is not faster compared to standard desktop tools. However, it is considered better because of the feeling of immersion and perception of collaboration with a co-worker.

\subsection{Discussion}\label{sec4_12}

In this paper, we stated that collaborative modeling in VR can now converge in speed to that of a standard desktop version and that VR is preferred by most developers in our experiment.

In addition, VR could gain importance within distributed teams, working remotely also due to pandemic situations, lockdowns, and restrictions. VR immerses people in collaborative modeling, and the change in work style would bring refreshment and possibly better results, as we mention in points 1 to 5 below. VR also makes sense by distributing models across interconnected layers or tables in its extended space.

After the result analysis and evaluation of the particular research hypotheses in the previous chapter 4.10, we can derive the answers to the research questions from the beginning of this study:

• RQ1. What is the impact of the created VR environment on the efficiency of collaborative, distributed software development?

Answer to RQ1: We have to conclude that VR tools converge to the efficiency of standard desktop tools. The results may be better if developers work longer with VR and get used to it, and headsets are updated as well. In the software domain, we need to improve the keyboard, voice input, OCR, etc. 

• RQ2. What is the impact of the created VR environment on the recall of design information during collaborative, distributed software development?

Answer to RQ2: The results of both experiments do not support our expectation that more intensive immersion in the virtual space with the model would improve the perception of its content and the retention of information.

• RQ3. What are the perceptions and preferences of users when using the VR-based compared to the non-VR software design environment?

Answer to RQ3: Most developers preferred to work in VR after the experiment ended. In future work, it would be interesting to let them work with the VR tool for a few months on a real project and evaluate afterwards their feelings, preferences and proposals about the environment settings, size and sensitivity of objects, operations of creating, editing, copy/paste, delete, zoom, hide, filter, and headsets calibration, etc.

It would be useful to take advantage of the developers' openness to such a change of interface as VR where they appreciate the change of working style and the novelty of the environment, but it would be necessary to very sensitively and carefully evaluate the emerging comments and suggestions and incorporate them into the next improved versions of the environment.

From the comparative analysis of the results, different properties of both versions, and conditions of both series of experiments, we can conclude as follows:

\begin{itemize}
    \item in the second version of VRsketch we have significantly refactored interface in Virtual Reality (e.g. improvement of VR keyboard, interaction with the whiteboards, selections of relationship attributes, speech recognition, teleporting, voice-call over tool, better avatars, see Chapter~\ref{sec3_4}) and therefore we got better times,
    \item in the second version of our experiments we divided participants into separate rooms (laboratories) and they were not satisfied with the collaboration in the desktop system (OctoUML) as much as with VRsketch. 
\end{itemize}

Although the task completion times in VR are still worse than in standard desktop software with a real keyboard and a mouse, there are obvious benefits from our experiences in experiments in 2020 and 2021/22:

\begin{enumerate}
    \item VR modeling offers positive changes in work style, replacing all-day sitting at the computer with movement and changing sitting position to help creativity and increase effectiveness and positive effect on thinking \cite{bib21, bib6},
    \item the collaboration with other human beings and co-workers is a more immersive experience in VR because we can see the movement of collaborator avatars, their gestures, where they are, and what they are observing or preparing to do,
    \item VR can enlarge the space for modeling to be more comprehensible for reading and more comfortable for editing,
    \item 3D space in VR is a better alternative to present the layers (using whiteboards, for example) for presenting particular components or interconnected distributed clusters,
    \item 3D space in VR clearly visualizes the model layer and source code layers behind it.
\end{enumerate}

Also, we obtained a neither positive nor a negative result: developers do not remember more information from work in VRsketch than they do from work in OctoUML, which means that the second research hypothesis was not confirmed.

\begin{figure}[h]
    \centering
    \includegraphics[width=0.3\textwidth]{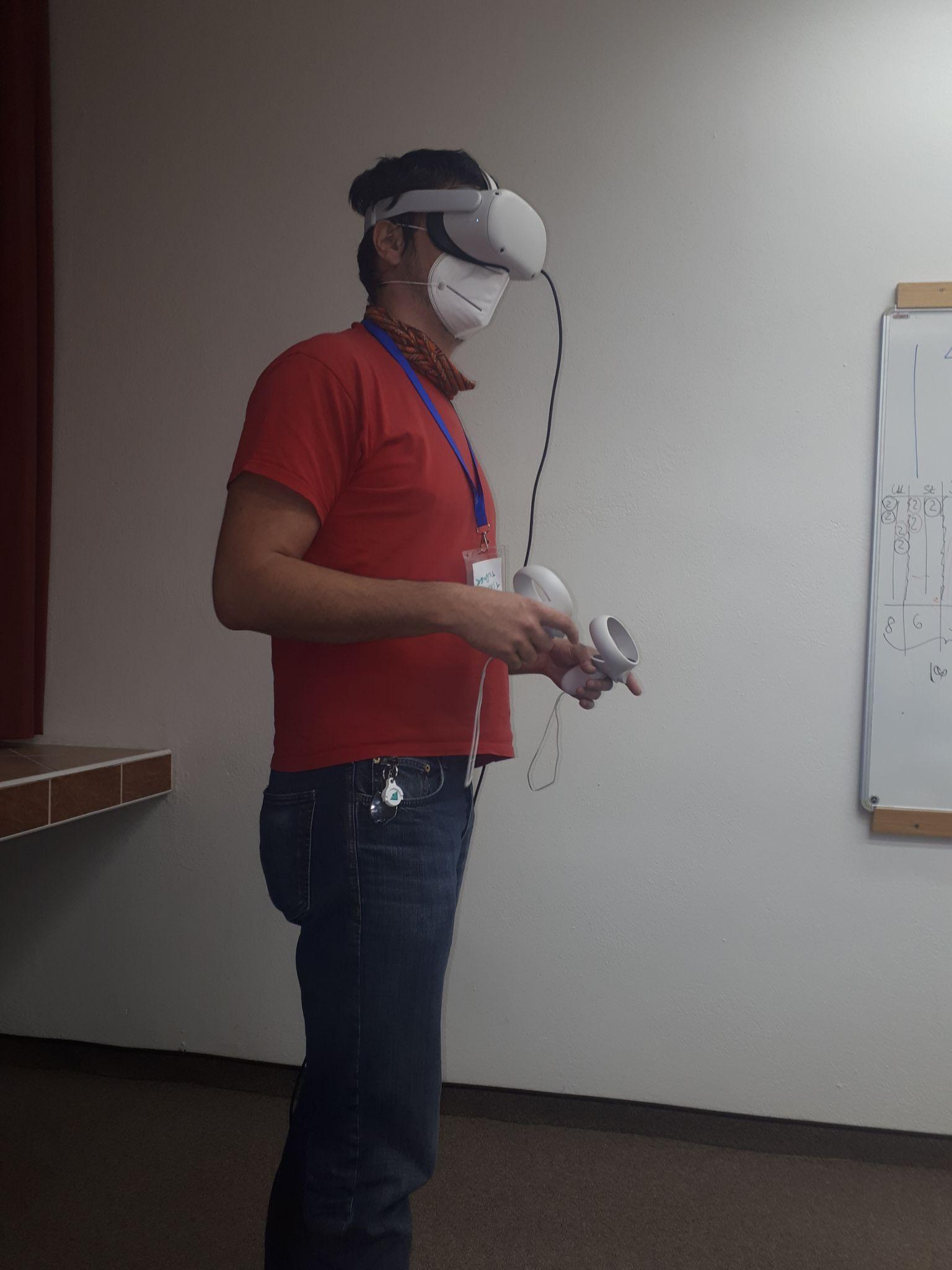}
    \caption{Second phase of experiments with the face masks}
    \label{fig:experiment_with_mask}
\end{figure}

On average, the times in OctoUML are still slightly better than in VRsketch, and we can improve the VRsketch times with a more in-depth training in VR, lasting several days focusing on a specific task. We are able to achieve even better times using alternative speech recognition without a face mask (mandatory due to the covid pandemic in the second series of experiments, as shown in Figure~\ref{fig:experiment_with_mask}).

We attempted to reduce the errors in the experiment by dividing it into three groups.

\begin{enumerate}
\item \textbf{Threats to Construct Validity}: In our efforts to design a suitable experiment, we aimed to create a correct procedure that allowed participants to learn both the VR and conventional desktop tools and become familiar with the tasks. However, we encountered a limitation concerning the time allocated for training with VR. The relatively short duration, lasting only a few minutes, might not have been sufficient to achieve comparable expertise when compared to using a traditional desktop system. 
Unfortunately, developers would not participate in an experiment in which they had to learn a new VR system for at least a day or ideally a week in a real project to become comfortable working in it naturally.
\item  \textbf{Threats to internal validity} (to ensure that the observed effect really does exist due to the test conditions \cite{BB}). We tried to map the focus and task content of the experiment to the use of such systems in the real world. We tried to avoid subjectivity in the design of the procedure and tasks, the content of the questionnaires and the assessment of individual variables to avoid errors and so on. We tried to reduce any bias of the experiment designers that a VR or desktop application would be better in specific tasks.
\item \textbf{Threats to external validity} (results are generalizable to other people and other situations \cite{BB}). We tried to recruit participants from different areas and assign them to the experiment randomly and without conscious selection, e.g. by experience, age, gender, etc. We created conditions of equal environment, time and material used for the experiment.
\end{enumerate}

We indirectly addressed some problem areas in our research, cybersickness \cite{X2, Y11}, and cognitive load \cite{Y6}. To track satisfaction with the tool, we utilized post-task questionnaires with SUS and NASA TLX ratings. The satisfaction responses also reflected the impact of these undesirable effects. Some participants rated the tool with lower satisfaction levels due to discomfort and dizziness, which cannot be entirely eliminated, even with collaborative modeling. While our main goal was to enhance collaborative work efficiency and intensity through VRsketch, we acknowledge that cybersickness and cognitive load are important considerations. We recognize that some developers may prefer the desktop application, and we do not expect them to work in VR all day. Instead, VRsketch aims to facilitate more effective and intensive collaboration, allowing collaborators to work together in a VR space, designing or reorganizing models using whiteboards and collaborative tools.

Batmaz et al. in \cite{ACM} identified that a change in target depth negatively affects virtual hand interaction in peri-personal space with single-focal displays. However, multifocal displays do not suffer from this issue. These results are consistent with the findings of the Barrera and Stuerzlinger \cite{88} for 3D TVs and Batmaz et al. \cite{99} for VR and AR headsets. Notably, they did not observe corresponding problems for lateral movements in \cite{ACM}. Furthermore, none of their experiment participants complained about eye strain.

We assume that collaborative modelling sessions in real-life scenarios would typically last for about 1-2 hours. Given the short-term use, the risk of VAC might be not significant: our experiments were not completed by just two people in v1 experiment from 152 participants (44 people in v1 experiment + 8 people in v2 experiment + 80 people in v3 experiment). In addition, our system prioritizes the lateral dimension over the depth dimension, utilizing flat whiteboards with UML models and excluding 3D geons and background.

This might not be a major problem even in future work in interconnected layers, as our experience with 3D UML suggests that we do not perceive all layers simultaneously. Instead, we tend to view them in a layer-interconnection-layer manner. Nevertheless, this aspect warrants more accurate monitoring, such as through the use of the Simulator Sickness Questionnaire \cite{sss}, rather than relying solely on indirect NASA TLX measurements. We might have better time results with a multifocal headset in the future.

One limitation of our project is that we used standard quality avatars and texts in VR. Our primary focus was on optimizing the keyboard, editor, and environment interface. While there are already sophisticated avatars \cite{Y10, X1} available in games and VR applications, there are also simpler alternatives \cite{Y3, Y8} for scenarios where the visual fidelity is not a critical advantage, and computing power needs to be prioritized for other essential functionalities. Our approach falls in the middle ground; we aimed for avatars resembling colleagues and providing insight into their actions, such as their gaze direction and current activities. This is complemented by tracking his work (VRsketch highlights the currently updated classes using a unique color assigned to specific developer). For future development, we envision the possibility of constructing avatars as point clouds \cite{Z} directly from photos or by capturing the developer's face through the computer's camera.

Why did our experiments yield better results in terms of efficiency, number of errors, and preferences than our colleagues at VmodlR \cite{bib37}?

It is not possible to rigidly compare particular results because we did not use the same tasks, participants, and base desktop standard 2D tools for comparison (we used OctoUML and they used Lucidchart). We did not know their research in 2020 (they published their tool and experiments in Models 2021), and they did not know our approach and results either. But maybe there is one objective reason for different results: the distinctions or inequality in the graphical user interface. We used 2D whiteboards in 3D virtual reality and not 3D boxes on green grass, which may have been more difficult to work with, although the engine inside both of our tools was comparable in terms of the libraries and environments used.

So far, we had no time or opportunity to test the effects of the infinite extent of the surfaces (whiteboards) in VR working on a large-scale project. In the future we will rely on simplifying the cognitive load and complexity by using the distribution of the project content into multiple layers – connected whiteboards that divide the project according to modules, use cases, etc. (e.g. also into a model layer and a source code layer).

After long discussions about the advantages and disadvantages of VR tools, we proceeded to perform the third short experiment, where an experienced professional software designer in a standard tool (OctoUML) confronted one of the main authors of VRsketch, who lost in the ratio of times $T_{VRsketch}$ vs $T_{OctoUML}$  3:33 vs 2:52 for the first task and 3:39 vs 2:54 for the second task.

\section{Conclusion and Future Work}\label{sec7}

The aim of this research was to support as well as improve the efficiency of distributed software design activities. We designed and evaluated a VR software design environment (called VRsketch) that supports the collaboration experience between individual and geographically distributed software developers. 

We conducted two experiments to evaluate the: (i) collaboration efficiency, (ii) recall of collaboration experience, and (iii) user preference as well as satisfaction when using the VR software design environment compared to a non-VR software design environment.

The results show that there is no significant difference in the efficiency and recall of design information when using the VR compared to the non-VR environment. After the first experiment, we had to improve the prototype of the VR environment to achieve better collaboration efficiency. To do so, we had to redesign the GUI of the prototype VRsketch to increase the convenience and efficiency to complete VR tasks in the second experiment. The results have improved. The difference between the execution times (i.e., efficiency) in VRsketch and the non-VR design collaboration environment was not statistically significant.

Based on our experience from the observations of the two experiments, we could improve voice input and hand gesture control with haptics and a possible addition of extended reality, where we could combine real keyboards and large comfortable visualizations, models or whiteboard in VR around a desktop computer or a laptop to improve collaboration efficiency.

Regarding user preference and satisfaction, we have found that developers are more satisfied with collaborating in a VR environment rather than in a standard desktop environment, thus it is important to further improve the prototype for a better support of collaborative software design in VR.

\subsection{Future Work}

Our prototype is open for adding more extensions in the future: for example, user customized real-world avatars and collaborative modeling of the other UML diagrams (Use Case Diagram, Sequence, Activity, and State Diagrams, etc.). We can create interconnected whiteboard layers in virtual reality for particular modules, clusters or element types (GUI, Business services, DB services) or for lean architecture with Data, Context, and Interaction in tiers \cite{bib24}.

We are designing a clustered shallow and a deep copy which will include copying of all dependencies and relations in a cloud of classes. Copied classes will act like clones, which means that every change in the original class (such as deleting/adding attributes, methods, relations) will be visible on the clone (copied class). This strategy can support pair-training of software modeling and the use of design patterns in large models for software houses.

\section{Acknowledgment}
We would like to thank Gratex International for providing the space and their developers for the experiments. We would also like to thank Faculty of Mathematics, Physics and
Informatics Comenius University Bratislava for providing the facilities and Dr. Pavel Petrovic for encouraging and supporting his students to participate in the experiments.

This work has been carried out in the framework of the TERAIS project, a Horizon-Widera-2021 program of the European Union, GA no. 101079338.

\section{Ethical Issues}
In this study, we have considered all the main ethical issues according to \cite{bib25}: informed consent, beneficence - no harm, respect for anonymity and confidentiality.

\bibliographystyle{elsarticle-num}

\end{document}